# Analysis of Volatility in Driving Regimes Extracted from Basic Safety Messages Transmitted Between Connected Vehicles

**Asad J. Khattak, Ph.D.**
Beaman Professor & Transportation Program Coordinator
University of Tennessee, Knoxville, TN 37996, USA
akhattak@utk.edu

**Behram Wali (Corresponding Author)**
Graduate Research Assistant, PhD Student
Civil and Environmental Engineering Department
University of Tennessee, Knoxville, TN 37996, USA
bwali@vols.utk.edu

**Abstract –** Driving volatility captures the extent of speed variations when a vehicle is being driven. Extreme longitudinal variations signify hard acceleration or braking. Warnings and alerts given to drivers can reduce such volatility potentially improving safety, energy use, and emissions. This study develops a fundamental understanding of instantaneous driving decisions, needed for hazard anticipation and notification systems, and distinguishes normal from anomalous driving. In this study, driving task is divided into distinct yet unobserved regimes. The research issue is to characterize and quantify these regimes in typical driving cycles and the associated volatility of each regime, explore when the regimes change and the key correlates associated with each regime. Using Basic Safety Message (BSM) data from the Safety Pilot Model Deployment in Ann Arbor, Michigan, two- and three-regime Dynamic Markov switching models are estimated for several trips undertaken on various roadway types. While thousands of instrumented vehicles with vehicle to vehicle (V2V) and vehicle to infrastructure (V2I) communication systems are being tested, nearly 1.4 million records of BSMs, from 184 trips undertaken by 71 instrumented vehicles are analyzed in this study. Then even more detailed analysis of 43 randomly chosen trips (N = 714,340 BSM records) that were undertaken on various roadway types is conducted. The results indicate that acceleration and deceleration are two distinct regimes, and as compared to acceleration, drivers decelerate at higher rates, and braking is significantly more volatile than acceleration. Different correlations of the two regimes with instantaneous driving contexts are explored. With a more generic three-regime model specification, the results reveal high-rate acceleration, high-rate deceleration, and cruise/constant as the three distinct regimes that characterize a typical driving cycle. Moreover, given in a high-rate regime, drivers' on-average tend to decelerate at a higher rate than their rate of acceleration. Importantly, compared to cruise/constant regime, drivers' instantaneous driving decisions are more volatile both in "high-rate" acceleration as well as "high-rate" deceleration regime. The study contributes to analyzing volatility in short-term driving decisions, and how changes in driving regimes can be mapped to a combination of local traffic states surrounding the vehicle.

## 1. Introduction

As a crucial part of technology driven progressive life, automobiles and transportation systems have continued to advance since its inception decades ago. The advent of rapid technological advancements in recent decades have established the elemental foundation for Cooperative Intelligent Transportation Systems (C-ITS), a.k.a. connected and automated vehicles. This said, equipping motor vehicles and transportation systems with wireless communication technologies in a bid to establish cooperative, well informed, and proactive transportation systems is expected to be the next frontier of transportation revolution (Lu et al., 2014, Fagnant and Kockelman, 2015). Specifically, connected and automated vehicle technologies refer to integrated systems that establish bidirectional wireless connectivity among vehicles itself (vehicle-to-vehicle V2V) and the infrastructure (vehicle-to-infrastructure V2I) to capture vehicle position, motion, vehicle maneuvering and instantaneous driving contexts[1] (Kamrani et al., 2017, US-DOT, 2016).

The generated large-scale integrated empirical data from connected and automated vehicles has significant potential in facilitating deeper understanding of instantaneous driving decisions[2]. Variations in driving with respect to the ecosystem of mapped local traffic states in close proximity surrounding the host vehicle can be explored. Important in this respect is the concept of "driving volatility" that captures the extent of variations in driving, especially hard accelerations/braking and jerky maneuvers, and frequent switching between different driving regimes[3] (Khattak et al., 2015, Liu and Khattak, 2016, Wang et al., 2015, Kamrani et al., 2017). However, a fundamental understanding of instantaneous driving decisions is needed for hazard anticipation and notification systems, and for distinguishing normal from anomalous driving. The research issue is to explore different regimes of typical driving behavior and how long they last and the key correlates associated with each regime.

As a part of U.S. Department of Transportation's (USDOT) Real-Time Data Capture and Management Program, Safety Pilot Model Deployment (SPMD) in Ann Arbor, Michigan features real-world demonstration of connected vehicle safety applications, technologies, and systems by hosting approximately 3,000 vehicles instrumented with V2V and V2I communication systems (Henclewood, 2014). Altogether, 75 miles of roadway in Ann Arbor, Michigan are instrumented with roadside equipment (RSE) that are capable of communicating with appropriately instrumented vehicles, and devices via advanced communication and sensor technologies such as dedicated short-range communications (DSRC) (Henclewood, 2014). Furthermore, data acquisition systems (DAS) are installed in vehicles to facilitate V2V and V2I infrastructure communications. The core output from DAS are Basic Safety Messages (BSM) that describe (frequency of 10 Hz) vehicle's instantaneous position (latitude, longitude, and elevation), motion (vehicle speed, longitudinal and lateral acceleration), vehicle maneuvering (acceleration pedal, brake pedal and cruise control) and instantaneous driving contexts (number of objects around host vehicle, distance to the closest object, and relative speed of the closest object) (Henclewood, 2014, Liu and Khattak, 2016). The availability of such large-scale high resolution data is successfully used for developing a basis for improved real-time alerts, warnings, and control assistance applications (Liu and Khattak, 2016, Kamrani et al., 2017).

---

[1] In this study, instantaneous driving contexts refer to the surroundings of host vehicle equipped with V2V and V2I technologies. An example can be how much a driver constrained is in terms of different objects surrounding the vehicle and the distance of the host vehicle to the surrounding objects.

[2] By instantaneous driving decisions, we mean the instantaneous decisions that driver may undertake to navigate the vehicle from one point to another. Such decisions may include decisions in longitudinal direction such as speeding, braking, high-rate acceleration, and/or high-rate deceleration, or in lateral direction such as lane change maneuvers. However, throughout the paper, we use the term "instantaneous driving decisions" to refer to driving decisions in longitudinal direction.

[3] In Economics literature, the key variable(s) that characterizes time-series system(s) occasionally exhibit dramatic breaks or abrupt changes in its behavior. The portions of data profile before and after the abrupt change are typically referred to "regimes" (Hamilton, 2010). In this paper, we refer to the abrupt changes that may be expected in a typical driving cycle as "driving regimes".

By using real-world large-scale data transmitted between connected vehicles and infrastructure, the present study creates new knowledge for connected vehicle technologies by explicitly investigating time-series instantaneous driving decisions (and the embedded regimes) of connected vehicle drivers at a detailed microscopic level, and mapping such decisions to instantaneous driving contexts. This analysis is important in sense that driving decisions (e.g., acceleration or deceleration decisions) primarily depend on surrounding traffic states (Åberg et al., 1997, Haglund and Åberg, 2000, Choudhury, 2007, Choudhury et al., 2010), and a detailed understanding of driving decisions can significantly help us with better anticipating hazardous situations and providing warnings and alerts to drivers.

## 2. Literature Review

A careful review of literature reflects the prompt response by government agencies, automotive industry and academia to such disruptive yet beneficial connected and autonomous vehicles innovation. Recently, the proceedings of 9th University Transportation Centers (UTC) Spotlight Conference by the Transportation Research Board (TRB) on connected and automated vehicles reflected the perspectives of several stakeholders in order to assemble a goal oriented road map to achieve maximum benefits from connected and automated vehicle technologies (Turnbull, 2016). Specifically, efficient and reliable transportation connectivity solutions are being explored for its applicability to address real world safety challenges (Kamrani et al., 2017, US-DOT, 2016, Fagnant and Kockelman, 2015, Hu et al., 2015, Khattak et al., 2015, Kim et al., 2007, Liu and Khattak, 2016), mobility problems (Zhu et al., 2009, Hu et al., 2015, Zhu and Ukkusuri, 2015, Weber, 2015, Koulakezian and Leon-Garcia, 2011, Zeng et al., 2012, Kianfar and Edara, 2013, Moylan and Skabardonis, 2015, Genders and Razavi, 2015), and environmental challenges (Wang et al., 2015, Liu et al., 2015b, Fagnant and Kockelman, 2015, Shin et al., 2015, GM, 2015, Weber, 2015, Zeng et al., 2012, Liu et al., 2016, Kamalanathsharma and Rakha, 2016). Such emerging applications together with connected vehicle infrastructure deployment strategies can address potential challenges related to operations and safety which can in turn benefit state and local transportation agencies (Hill and Garrett, 2011).

Connected and automated vehicle solutions can potentially help in addressing transportation challenges by primarily targeting the human factor involved in surface transportation. In special relevance to transportation safety solutions, several studies have focused on monitoring driving behavior to develop cooperative collision warning systems (Sengupta et al., 2007, Yang et al., 2000, Chrysler et al., 2015, Goodall et al., 2016, Osman et al., 2015, Doecke et al., 2015, Lee et al., 2002, Lee et al., 2004, Abe and Richardson, 2006, Naseri et al., 2015). By carefully characterizing driving behavior, the afore-mentioned studies contributed by developing effective collision warning systems and documented the potential of connected vehicle technologies in addressing major transportation safety challenges (Chrysler et al., 2015, Goodall et al., 2016, Osman et al., 2015, Doecke et al., 2015). However, the previous studies either utilized driving simulator/algorithm developments or localized closed course experiments, which may not cover different driving contexts/conditions. Moreover, the key to success of connected vehicle technologies rely on how well and effective connectivity of vehicles and/or infrastructure can perform in real life situations. Important in this regard are the recent innovations that enable realization of V2V and V2I technologies such as DSRC, Wi-Fi, Bluetooth, and cellular networks (Cheng et al., 2007, Chou et al., 2009, Sugiura and Dermawan, 2005).

Towards this end, recent studies utilized large scale behavioral data integrated with sensor technologies to introduce the concept of "driving volatility", which can be regarded as a measure of driving practice for characterizing instantaneous driving decisions and more specifically extreme driving behaviors (Wang et al., 2015, Liu et al., 2015b). The studies by (Wang et al., 2015) and (Liu et al., 2015b) investigated relationships between driving volatility (for each trip) and factors such as driver demographics, trip related factors (purpose, duration) and detailed vehicle characteristics such as body type, fuel type, transmission and power train (Wang et al., 2015, Liu et al., 2015b). Collectively, the potential of individual level driving volatility in developing advanced traveler information systems, driving feedback devices, and alternative fuel vehicle purchase frameworks for consumers was documented (Wang et al., 2015, Liu et al., 2015b). Likewise, (Noble et al., 2014) utilized naturalistic driving data collected through the Strategic Highway Research Program 2 for developing a vehicle to infrastructure (V2I) warning algorithm. Specifically, in realistic driving behavior context, (Choudhury, 2007) and (Choudhury et al., 2010) focused on developing framework for "more realistic" driving lane changing and freeway merging behavior models that accounted for "unobserved driving plans" behind the observed driving decisions (Choudhury, 2007, Choudhury et al., 2010). Among other innovative techniques, Hidden Markov Models were introduced to account for "regime-dependence" in driving decisions in congested and freeway merging scenarios, where the current driving plan depended on all previous actions (Choudhury, 2007, Choudhury et al., 2010). In addition to simulation validations, empirical vehicle trajectory data was used to justify the use of regime-dependent plans in microscopic traffic simulator environment (Choudhury, 2007). While afore-mentioned studies provided valuable information about driving actions (Noble et al., 2014) and extreme driving events (Wang et al., 2015, Liu et al., 2015b), such extreme events could not be mapped to local traffic conditions due to unavailability of data. Similarly, the study by (Choudhury, 2007) focused on lane changing and freeway merging driving decisions, and not micro-level instantaneous driving decisions and the impact of local traffic conditions on instantaneous driving decisions.

SPMD provides an exciting opportunity by using state-of-the-art technologies to generate Basic Safety Messages (BSMs) that describe vehicle's instantaneous position, vehicle maneuvering, and instantaneous driving contexts (Henclewood, 2014). In special relevance to current study, study by (Liu and Khattak, 2016) extracted critical information from raw BSMs that captured trip level extreme driving events. An understanding of occurrence of extreme driving events was sought by identifying its correlates such as trip attributes, vehicle maneuvering and driving context for successful generation of real-time improved alerts, warnings, and control assistance systems (Liu and Khattak, 2016). While the study by (Liu and Khattak, 2016) utilized large-scale BSM data sent and received by vehicles and roadside equipment, the study primarily focused on conceptualizing trip-level extreme driving events (based on specific thresholds) and did not explore the instantaneous driving actions (within the trip) and its associations with instantaneous driving contexts that are taken along a specific trip.

### 2.1. Research Objective

Given the prevalent gap in connected vehicle literature, the present study builds upon the existing body of connected vehicle knowledge by focusing on, 1) categorizing time-series based driving tasks[4] into different regimes using information contained in BSMs; 2) categorizing the volatility in each regime and the average duration of each regime, and 3) Identifying the correlates that can be associated with drivers' tendency to stay in a specific regime and/or to switch between different regimes. By doing so, a fundamental understanding of instantaneous short-term driving decisions is sought (with respect to different roadway types) and how can we map time-series instantaneous driving behavior to a combination of local traffic states such as instantaneous driving contexts. Given the temporal dependency in instantaneous driving decisions, the current study methodologically contributes by introducing rigorous dynamic Markov switching models for conceptualizing micro-level driving behavior into different regimes, while mapping correlates to each regime. To the best of our knowledge, for a deeper understanding of instantaneous driving decisions, such time-series models together with utilization of large-scale real-world connected vehicle data have not been used.

## 3. Methodology

### 3.1. Conceptual Framework

A key objective of this study is to explore volatility in driving behavior by applying appropriate analytic tools to identify the correlates of instantaneous driving decisions. At a basic level, instantaneous driving decisions can be categorized into at least two regimes, and drivers can switch between these regimes over time. The two regimes/states are unobserved yet distinct, in the sense that in the different regimes, instantaneous driving decision data are generated by separate continuous processes (Hamilton, 1989). By separate continuous processes we mean that data generation in two regimes along a trip can be developed by different effects of instantaneous driving contexts and assuming a time-constant association/effect across a trip irrespective of different regimes may overlay the true data generation process[5].

Therefore, for simplicity and illustration, we first categorize instantaneous short-term driving performance into two regimes. While incorporation of additional regimes is conceptually valid and theoretically possible, doing so significantly complicates the modeling framework due to computational tractability and regime identification issues (discussed later in detail). This is evident from the literature where models with more than two regimes are not common and different time-dependent regime varying processes (such as traffic crashes, economic, or financial data) are usually modelled as a two-regime processes, e.g. (Hamilton, 1989, Hamilton, 1994, Malyshkina and Mannering, 2009, Malyshkina et al., 2009, Hansen, 1992, Kim and Nelson, 1999) and the references therein. Nonetheless, not in transportation field though, very few studies have also considered three-regime models for modeling different financial and economic time-series datasets (Hardy, 2001, Kim et al., 2008).

---

[4] In this paper, the term "driving task" refers to the combination of instantaneous driving decisions that driver may take in the longitudinal direction along an entire trip. Depending on the context, we use the term driving task interchangeably with the term "driving cycle".

[5] There can be several reasons to anticipate existence of two regimes. Depending on several factors, instantaneous driving decisions (magnitude and directions of longitudinal accelerations) can vary significantly across the entire trip. Thus, under potentially different conditions (i.e. different instantaneous driving contexts), drivers may respond differently to staying in the same regime or switching to a different regime.

On the other hand, real-world driving is a complex task and we can anticipate existence of more than two regimes, say three regimes in a typical driving cycle. Thus, as pointed out by the reviewers too, it is plausible to start with a more generic model specification that may capture common driving regimes, and thus can help in extracting important information related to instantaneous driving decisions embedded in real-world connected vehicle data[6]. Having said this, we thoroughly investigate real-world instantaneous driving decisions in connected vehicle environment based on two and three regime dynamic Markov switching models.

Next, we investigate associations of instantaneous driving decisions with critical correlates (available in the data) related to instantaneous driving context such as the number of objects around the host vehicle and distance to the closest object. By doing so, a fundamental understanding of instantaneous short-term driving decisions is sought (with respect to different roadway types) and how can we map time-series instantaneous driving behavior, especially driving volatility to a combination of local traffic states such as instantaneous driving contexts. This is important in the sense that instantaneous driving contexts, at least at a basic level, can be represented by surrounding vehicles around the host vehicle which may constrain movement and/or motivate driver to get out of congested situation. Assuming (for now) that the driver's tendency is to get out of congested situations, how the driver actually maneuvers the car is an important question which is likely to have important safety (among others) implications (Liu and Khattak, 2016). Is there frequent switching from acceleration to braking and vice versa? These behaviors are perhaps more dangerous, compared with other behaviors such as constant speed (Liu and Khattak, 2016).

   As instantaneous driving behavior (across an entire trip) is a time-varying process, we use a Markov regime switching dynamic regression framework that assumes Markov switching (over time) between two and three (unobserved) regimes in a typical driving cycle. Note that the regime switching can be based on change in measures of central tendency (averages) and/or dispersion (variance). Having said this, conceptualizing the driving task into two (or three) different regimes can potentially account for existence of several unobserved factors that may be associated with driving performance envelope (Hamilton, 1989). Markov switching models thus can treat driving behaviors in an intuitive manner. As a matter of fact, two-regime Markov switching models are used successfully in solving problems related to traffic safety, for exhaustive applications of Markov switching regressions in safety area, interested readers are referred to (Malyshkina et al., 2009, Xiong et al., 2014, Malyshkina and Mannering, 2009).

---

[6] We sincerely thank the two reviewers for suggesting investigation of more than two-regimes in a typical driving cycle. Doing so came at a cost of losing some data (discussed later in detail), nonetheless, exploration of three regime instantaneous driving behavior models helped us in extracting meaningful information from the data which was otherwise not possible from the two-regime specification.

Figure 1 presents the hypothesized behavior during a general trip where "1" refers to regime 1; "2" refers to regime 2 and P(1-1) indicates the probability that a driver in regime 1 at current time will continue in regime 1 during the next time period. Figure 1 also illustrates the time-series framework as a Markov regime switching dynamic regression. Assume that a driver is currently (at time instant t = -1 seconds) in regime 1; the driver at next instant of time (t = 0 second) can either decide to remain in regime 1 or to switch to regime 2, given the effects of correlates, i.e. instantaneous driving contexts. If the driver is in regime 2 (or vice versa) at t = 0 second, the challenge is to predict driver action at next instant of time (indicated by t = 1 second) given the effects of associated covariates.

Following similar concept, Figure 2 presents a three-regime typical driving cycle based on Markov Switching dynamic regression framework where "1" refers to regime 1; "2" refers to regime 2; and "3" refers to regime 3. If a driver is currently (at time instant t = -1 seconds) in regime 1; the driver at next instant of time (t = 0 second) can either decide to remain in regime 1 or to switch to regime 2 or regime 3, given the effects of correlates, i.e. instantaneous driving contexts. If the driver is in regime 2 (or vice versa) at t = 0 second, the challenge is to predict driver action (to stay in regime 2, or to switch to regime 1 or 3) at next instant of time (indicated by t = 1 second) given the effects of associated covariates.

**(PLACE FIGURE 1 ABOUT HERE)**
**(PLACE FIGURE 2 ABOUT HERE)**

With the empirical framework of two and three regimes Markov Switching dynamic regression models, the research questions are:
- What are these regimes in typical driving cycle?
- How much is the volatility each regime?
- When do the regimes change or how long they last?
- Are driver decisions consistent across different trips undertaken by different drivers? Precisely, while allowing for differential effects of key correlates across two and three regimes, are the correlations constant across the regimes?

Finally, the proposed methodology has the potential to probabilistically predict a driving regime at a specific instant of time while allowing for the effects of instantaneous driving contexts. This is important in the sense that a change from one regime to another is not perfectly deterministic due to several unobserved factors. Thus, a time-series model should account for the probabilistic nature of the process. The proposed conceptual framework is focused on answering the afore-mentioned critical questions. A detailed description of formulating the given problem in a mathematical framework is presented in later section.

### 3.2. Markov-switching dynamic (abrupt-change) regression models

#### 3.2.1. Two-Regime Dynamic Markov-switching regression models:

Markov switching models were recently introduced in traffic crash modeling for addressing different important issues related to traffic safety, for exhaustive applications of two-regime Markov switching regressions in safety area, interested readers are referred to (Malyshkina et al., 2009, Xiong et al., 2014, Malyshkina and Mannering, 2009). As instantaneous driving behavior (across an entire trip) is a time-varying process, we use a Markov regime switching dynamic regression (MSDR) framework that assumes Markov switching (over time) between two (unobserved) regimes[7], in this case regime 1 and regime 2 for two-regime model. Consider the evolution of driving behavior "$y_t$", where t = 1, 2, .....,T (i.e. the entire duration of the trip) that is particularly characterized by two regimes/states:

Regime 1: $y_t = \mu_1 + \emptyset y_{t-1} + \varepsilon_t$ (1)
Regime 2: $y_t = \mu_2 + \emptyset y_{t-1} + \varepsilon_t$ (2)

Where: $\mu_1$ and $\mu_2$ are the intercept terms in regime 1 and regime 2 respectively; $\emptyset$ is the Autoregressive parameter; and $\varepsilon_t$ is the white noise with variance $\sigma^2$. The two regime model abrupt shifts in the intercept term (Hamilton, 1994). At times, if the timing of the switching is known to the analyst, the above models (Equation 1 and 2) can be expressed as:

$y_t = s_t\mu_1 + (1 - s_t)\mu_2 + \emptyset y_{t-1} + \varepsilon_t$ (3)

Where: $s_t$ is 1 if the process (driving behavior cycle) is in regime 1 and 2 if in regime 2. Empirically, the model in Equation 3 can be conceptualized as regression with dummy variables and can be estimated with ordinary least squares regression (Hamilton, 1994). However, in the case under consideration, we never know in which regime the process is at current time, or indirectly $s_t$ is unobserved[8]. This said, Markov-switching regression framework specifies that the unobserved $s_t$ follows a Markov chain.

Note that the transition of driving cycle between two regimes can either be abrupt-change (dynamic Markov switching specification) or gradual adjustment (Autoregressive Markov Switching specification) after the process changes regime. However, in our case, due to the high resolution (frequency of 10 Hz) of instantaneous driving behavior data (dependent variable), we allow the driving cycle for a specific trip to switch between two regimes abruptly and not with gradual adjustment, thus called Markov Switching Dynamic Regression (MSDR) (Hamilton, 1994). This alternatively suggests the autoregressive term "$\emptyset$" in equation 1 and 2 equals zero. Thus, in the simplest case, we can express the framework as regime-dependent abrupt-change intercept term for k regimes (in our case k = 2) as:

$y_t = \mu_{s_t} + \varepsilon_t$ (4)

Where: $\mu_{s_t} = \mu_1$ when $s_t$ =1 (i.e. regime 1) and $\mu_{s_t} = \mu_2$ when $s_t$ =2 (i.e. regime 2) and $\varepsilon_t$ is the white noise with variance $\sigma^2$. In the simplest case, with switching in variance term[9] "$\sigma^2$" and no explanatory variables, six parameters $\mu_1, \mu_2, \sigma_1{}^2, \sigma_2{}^2, p_{1\rightarrow2}, p_{2\rightarrow1}$ are estimated.

---

[7] The three-regime MSDR framework is explained later in this section.
[8] It is important to note that the dependent variable (instantaneous driving decisions in longitudinal direction) is observed, but the regimes ($s_t$) are not observed. That is, we as analysts do not know a-priori what specifically the two-regimes are that characterize a typical driving cycle. We explain this in detail in the results section.
[9] In addition to switching of intercept term, variances can be regime-dependent (separate variance for two regimes) or regime independent (single variance for the entire process). The decision to allow switching in variance terms can be based on empirical and/or theoretical evidence. In addition to empirical justification from data, we posit that the two unobserved regimes are two distinct components of driving behavior and the variance in the evolution of the two regimes can be significantly different from each other. Thus, constraining the variance term to be regime-independent can potentially hide (as we will show) the true information embedded in data generation process.

Furthermore, the conditional density of driving cycle $y_t$ is characterized by a first order two-state Markov process as:

$$f(y_t|s_t = i, y_{t-1}; \boldsymbol{\theta}) \tag{5}$$

Where $\boldsymbol{\theta}$ is a vector of parameters i.e. in simplest case with only intercept terms and regime-specific variances, $\boldsymbol{\theta} = [\mu_1, \mu_2, \sigma_1{}^2, \sigma_2{}^2, p_{1\to2}, p_{2\to1}]$. For two regimes, there are two conditional densities, and thus estimation of parameter vector $\boldsymbol{\theta}$ is performed by updating the conditional likelihood using nonlinear filter (Hamilton, 1994), as opposed to linear updates by (Harvey, 1990). With a vector of set of explanatory variables "B" along with switching intercepts, the general specification of MSDR can be written as (Hamilton, 1989):

$$y_t = \mu_{s_t} + X_t \propto +Z_t \beta_{s_t} + \varepsilon_t \tag{6}$$

Where: $y_t$ is the dependent variable, $\mu_{s_t}$ is the regime-dependent intercept term, $X_t$ is a vector of exogenous variables with regime-independent coefficients $\propto$, $Z_t$ is a vector of exogenous variables with regime-dependent coefficients $\beta_{s_t}$, and $\varepsilon_t$ is independent and identically distributed (i.i.d.) normal error with mean 0 and regime-dependent variance $\sigma_{St}{}^2$. In Equation 6, as the two regime variables $s_t$ are unobservable, the vector of estimable parameters for Equation 6 shall include $\boldsymbol{\theta} = [\mu_1, \mu_2, \sigma_1{}^2, \sigma_2{}^2, p_{1\to2}, p_{2\to1}]$ in addition to parameter estimates for regime-dependent and regime-independent explanatory variables[10].

### 3.2.2. *Three-Regime Dynamic Markov-switching regression models:*
The modeling framework can now be extended to a three-regime specification. Consider the evolution of driving behavior "$y_t$", where t = 1, 2, …..,T (i.e. the entire duration of the trip) that is particularly characterized by three unobserved regimes/states:

$$y_t = \tau_{s_t} + \varepsilon_t \tag{7}$$

Where:

$$\tau_{s_t} = \begin{cases} \tau_1 \; if \; s_t = 1 \; (\text{regime 1}) \\ \tau_2 \; if \; s_t = 2 \; (\text{regime 2}) \\ \tau_3 \; if \; s_t = 3 \; (\text{regime 3}) \end{cases} \tag{8}$$

And, $\varepsilon_t$ is the normally distributed white noise with mean 0 and variance $\sigma_{s_t}^2$, $s_t = $ (Åberg et al., 1997) is an unobservable state variable governed by a first-order Markov chain. In the simplest case, with switching in variance term "$\sigma^2$" and no explanatory variables, the parameter vector $\boldsymbol{\theta} = [\mu_1, \mu_2, \mu_3, \sigma_1{}^2, \sigma_2{}^2, \sigma_3{}^2, p_{1\to1}, p_{1\to2}, p_{2\to1}, p_{2\to2}, p_{3\to1}, p_{3\to2}]$, i.e. twelve parameters are estimated. Similar to the two-regime models, the three conditional densities (for three regions) associated with estimation of parameter vector $\boldsymbol{\theta}$ is performed by updating the conditional likelihood using nonlinear filter (Hamilton, 1994).

---

[10] In our case, we posit that the effects of explanatory variables (i.e. number of objects around host vehicle and distance to closest object) can be different with respect to two regimes. Thus, $X_t$ (vector of regime independent exogenous variables) is zero. As a result, the vector of estimable parameters for Equation 6 is $\boldsymbol{\theta} = [\mu_1, \mu_2, \sigma_1{}^2, \sigma_2{}^2, p_{1\to2}, p_{2\to1}, \beta_{s_t=1}, \beta_{s_t=2}]$, where $\beta_{s_t=1}, \beta_{s_t=2}$ are regime dependent vectors of estimable parameters for exogenous variables.

With a vector of set of exogenous explanatory variables "**W**" along with regime-dependent intercepts and variances, the general specification of a three-regime MSDR can be written as (Hamilton, 1989):

$$y_t = \tau_{s_t} + X_t\delta + Z_t\gamma_{s_t} + \varepsilon_t \tag{9}$$

Where: $y_t$ is the dependent variable, $\tau_{s_t}$ is the regime-dependent intercept term, $X_t$ is a vector of exogenous variables with regime-independent coefficients $\delta$, $Z_t$ is a vector of exogenous variables with regime-dependent coefficients $\gamma_{s_t}$, and $\varepsilon_t$ is independent and identically distributed (i.i.d.) normal error with mean 0 and regime-dependent variance $\sigma_{St}^2$. Given the inclusion of regime-dependent exogenous explanatory variables, the estimable parameter vector $\boldsymbol{\theta}$ is now expanded in Equation 9[11].

### 3.3. Markov chains

A discrete time Markov chain (DTMC) is assumed during switching mechanism of driving cycle between two regimes i.e. the probability distribution of $s_{t+1}$ depends only on current regime $s_t$ and not on the previous evolution of driving behavior[12] i.e. $s_{t-1}, s_{t-2}, \ldots..$ (Tauchen, 1986). This is commonly referred to a two-state Markov chain and is fairly a standard in applications of Markov Switching models (Hamilton, 1994, Malyshkina et al., 2009, Hamilton, 2010, Xiong et al., 2014). Higher order Markov chains where the realization of the future state may depend on current state and previous history brings in high complications to the model estimation process (Kim et al., 2008), and are thus not common in Markov switching applications[13]. Also, the first-order Markov chain seems a natural and intuitive starting point and, as mentioned in (Hamilton, 2010), is clearly preferable to acting as if the shift from regime 1 to 2 (or vice versa) was a perfectly deterministic event. Permanence, if any, of the shift between the regimes would be represented by $p_{2\to2}$ (in two regime case) equal 1, and any intra-regime probability of less than one (as we will see later) would indicate lack of permanence which the Markov formulation accommodates. Furthermore, if the regime change in instantaneous driving decisions reflects a change in instantaneous driving contexts, the prudent hypothesis would seem to be to allow the possibility for the regime to change back again when instantaneous driving context changes, and this suggests that $p_{2\to2} < 1$ is a more natural formulation for thinking about the regime changes than the deterministic $p_{2\to2} = 0$ (Hamilton, 2010, Kim et al., 2008). Having said this, assuming $s_t$ to be an irreducible and aperiodic Markov chain originating from its ergodic distribution $\pi = (\pi_1, \ldots \ldots, \pi_k)$, the probability that $s_t$ belongs to, $j \in (1,2)$ (where 1, 2 refers to regime 1 and 2) for two regime model and $j \in (1,2,3)$ (where 1, 2, and 3 refers to regime 1, 2, and 3) in three regime model depends on the most recent realization of driving behavior, $s_{t-1}$, and thus can be formulated as (Hamilton, 1994):

$$\Pr(s_t = j | s_{t-1} = i) = p_{ij} \tag{10}$$

---

[11] The parameter vector $\boldsymbol{\theta}$ for the three-regime MSDR framework has at least 15 parameters to be estimated, i.e. $\boldsymbol{\theta} = [\mu_1, \mu_2, \mu_3, \gamma_{s_t=1}, \gamma_{s_t=2}, \gamma_{s_t=3}, \sigma_1{}^2, \sigma_2{}^2, \sigma_3{}^2, p_{1\to1}, p_{1\to2}, p_{2\to1}, p_{2\to2}, p_{3\to1}, p_{3\to2}]$, where $\gamma_{s_t=1}, \gamma_{s_t=2}, \gamma_{s_t=3}$ are regime dependent vectors of estimable parameters for exogenous variables in the three-regime MSDR model.

[12] Another option can be to specify the models in continuous time. However, the advantage of DTMCs is that they have a mathematically easy formal description. A concern, however, can be that modeling continuous process is hard using a time-discrete paradigm. In other words, a uniform step must be artificially introduced, which will always result in errors and abstraction. However, in our case, we are not artificially introducing a time-step. Despite that driving cycle is a continuous process, we observe the driving decisions at discrete time intervals (t = 1, 2, 3, and so on.). Due to the very high data resolution of SPMD connected vehicle data, it is unlikely that drivers will make instantaneous driving decisions and perform frequent regime switching within one second. Also, the basic formulation of Markov property shows that observing a continuous-time Markov chain at regular time intervals gives a discrete-time Markov chain.

[13] An alternative and indirect way of extending the first-order Markov chain property can be to formulate a model specification where the evolution of response outcome may depend on the value of switching mean at its current state and lagged value, and this in turn will lead to four conditional densities where the new state variable is a four-state Markov chain. This specification is mathematically equivalent to Markov Switching Autoregressive framework as shown in Equations 1 and 2 and is typically used to model low frequency data (Hamilton, 1994, Kim, 1994). Keeping in view the extant literature, Markov switching dynamic regressions are used in the current study given the high resolution of instantaneous driving data (Stata, 2016, Hamilton, 2010).

Thus, all possible transitions from one regime to another, in a two-regime model, can be collected in $2 \times 2$ transition matrix while governing the evolution of Markov chain as:

$$\boldsymbol{P} = \begin{bmatrix} p_{1\to1} & p_{1\to2} \\ p_{2\to1} & p_{2\to2} \end{bmatrix} \tag{11}$$

While, the transition probabilities of switching from one regime to another in a three-regime model can be collected in a $3 \times 3$ transition probability matrix as:

$$\boldsymbol{P} = \begin{bmatrix} p_{1\to1} & p_{1\to2} & p_{1\to3} \\ p_{2\to1} & p_{2\to2} & p_{2\to3} \\ p_{3\to1} & p_{3\to2} & p_{3\to3} \end{bmatrix} \tag{12}$$

### 3.4.Likelihood function with latent states/regimes

Using the Markov chain property, the conditional density of $y_t$ can be formulated using Equation 5 for two or three regime models. However, in order to obtain marginal density of $y_t$, we weigh the conditional densities (one for each regime) by their respective probabilities, as explained in (Hamilton, 1994, Goldfeld and Quandt, 1973, Frühwirth-Schnatter, 2006):

$$f(y_t | \boldsymbol{\theta}) = \sum_{i=1}^{k} f(y_t | s_t = i, y_{t-1}; \boldsymbol{\theta}) \Pr(s_t = i, \boldsymbol{\theta}) \tag{13}$$

Over here, let us introduce a $k \times 1$ vector of conditional densities as:

$$\forall_t = \begin{bmatrix} f(y_t | s_t = 1; y_{t-1}; \boldsymbol{\theta} \\ f(y_t | s_t = 2; y_{t-1}; \boldsymbol{\theta} \\ \vdots \\ f(y_t | s_t = k; y_{t-1}; \boldsymbol{\theta} \end{bmatrix} \tag{14}$$

Where: k is number of regimes respectively. To construct the final likelihood function, the probability that $s_t$ takes on specific value (either 1 or 2 for a two-regime model or 1, 2, or 3 for a three-regime model) using the data through time "t" and model parameters $\boldsymbol{\theta}$ should be estimated. While utilizing the data until time "t", let $\Pr(s_t = i; y_t; \boldsymbol{\theta})$ denote the conditional probability of observing $s_t = 1$, then the resulting likelihood is:

$$\Pr(s_t = i; y_t; \boldsymbol{\theta}) = \frac{f(y_t | s_t = i, y_{t-1}; \boldsymbol{\theta}) Pr(s_t = i; y_{t-1}; \boldsymbol{\theta})}{f(y_t | y_{t-1}; \boldsymbol{\theta})} \tag{15}$$

The likelihood can then be estimated through iterating Equation 16 and 17 as[14]:

$$\aleph_{t|t} = \frac{\aleph_{t|t} * \forall_t}{1'(\aleph_{t|t-1} * \forall_t)} \tag{16}$$

$$\aleph_{t+1|t} = P\aleph_{t|t} \tag{17}$$

Where 1 is $k \times 1$ vector of constants i.e. 1s. The reduced likelihood representation is thus obtained as[15]:

$$L(\theta) = \sum_{t=1}^{T} log f(y_t | y_{t-1}; \theta) \tag{18}$$

Where:

$$f(y_t | y_{t-1}; \theta) = 1'(\aleph_{t|t-1} * \forall_t)$$

### 3.5. Predictions/regime prediction

To be able to predict the unconditional probability of a driving cycle in a specific regime at time "t", we use conditional transition probabilities and the Markov structure of the model. Specifically, the log-likelihood function has a recursive structure (Frühwirth-Schnatter, 2006) that initiates from the unconditional state probabilities $\aleph_{1|0}$. Thus, the unconditional probabilities are estimated as:

$$\pi = (\boldsymbol{A}'\boldsymbol{A})^{-1}\boldsymbol{A}' e_{k+1} \tag{19}$$

Where A is $(k + 1) \times k$ matrix formulated as:

$$A = \begin{bmatrix} \boldsymbol{I_k} - P \\ 1' \end{bmatrix} \tag{20}$$

---

[14] To achieve final likelihood function, we transform conditional probabilities for two regimes i.e. $\Pr(s_t = i; y_t; \boldsymbol{\theta})$ and $\Pr(s_{t-1} = i; y_t; \boldsymbol{\theta})$ to $k \times 1$ vector as $\aleph_{t|t}$ and $\aleph_{t|t-1}$ respectively.

[15] Characterization of maximum likelihood estimates has been performed through implementation of Expectation Maximization (EM) algorithm (Dempster et al., 1977). Due to the nonlinear equation structure for estimating parameter vector $\boldsymbol{\theta}$, it is practically not possible to solve them analytically, and as such, iterative algorithm is used to finding the maximum likelihood estimates. Each iteration of this algorithm consists of two simple steps: An E-step, in which a conditional expectation is calculated over a pre-defined density surface, and an M-step, where the conditional expectation is maximized. For a detailed discussion about EM algorithm in context of aperiodic ergodic Markov chains, interested readers are referred to (Hamilton, 1994).

And $I_k$ denotes $k \times k$ identity matrix, and $e_k$ denotes kth column of $\boldsymbol{I_k}$ respectively.

## 4. Data Description – Data Acquisition Systems

The data were extracted from the Data Acquisition System (DAS), which was part of Safety Pilot Model Deployment (SPMD) in Ann Arbor, Michigan. The key objectives of SPMD include evaluation of how drivers adapted to the utilization of connected vehicle technology, providing opportunity to explore real-world effectiveness of connected vehicle safety applications in multi-modal driving conditions (Henclewood, 2014). This study focuses on using the SPMD large-scale connected vehicle sanitized mobility data to understand instantaneous driver decisions in a broader ecosystem of instrumented vehicles and infrastructure on different roadway functional classifications.

As part of DAS, BSMs contain instantaneous (frequency of 10 Hz) information packets describing host vehicle's motion and location information, including vehicle performance (speed and acceleration), vehicle operation (brake and accelerator pedal application), and instantaneous driving contexts (number of objects around host vehicle and distance to the closest object) respectively (Henclewood, 2014). This information is stored in BSMs that are instantaneously sent and received by instrumented vehicles and roadside equipment (Henclewood, 2014). Table 1 summarizes the detailed description of key data variables whereas detailed description of all other data sources is available in SPMD Data Handbook (Henclewood, 2014). One-day sample data (04/11/2013) has been used for this study which contains approximately 1.4 million records (1,399,084) of basic safety messages, from 184 trips undertaken by 71 instrumented vehicles. Specifically, the sum of all trip durations is approximately 38.8 hours, whereas the average duration per trip is 12 minutes respectively. From roadway type stand-point, the overall trips are undertaken on combination of freeways, state routes, and local routes respectively.

### (PLACE TABLE 1 ABOUT HERE)

For this study, a probability based random-sampling procedure is conducted to randomly select 43 trips (out of 184 trips) for further analyses[16]. In the probability based simple random-sampling, random number generator (RNG) was used to generate unique indexes (ranging between 0 and 1) for each of the 184 trips and equal probability was assigned to each of the trip (i.e. probability of selecting each trip was same across the data matrix). Next, a sample of 43 trips is randomly extracted from the original data matrix (containing 184 trips) without replacement.

To facilitate more meaningful analysis, the entire vehicle trajectories for 43 randomly selected trips were visualized in Google Earth to identify the roadway functional classification on which the trips were undertaken. As such, significant efforts went into classifying the trips with respect to roadway type. For the sampled 43 trips, four trips are undertaken on freeway and state routes, 2 trips on US state routes, 14 trips on freeways, 18 trips on local roads, and 5 trips on state and local routes. Altogether, the 43 trips are undertaken by 34 vehicles whereas few vehicles undertook two or more than two trips.

---

[16] A total of 43 randomly selected trips were categorized and modeled at the microscopic level in this study. Analyzing the entire database of 184 trips was not done since it would be very labor intensive (in terms of categorizing) and computationally burdensome (in terms of modeling). Also, it is important to note that the 43 randomly chosen trips account for 52% of the total one-day BSM sample (714,340 BSM packets out of 1,399,084 packets).

The connected vehicle data used in this study are reliable and was error-checked. We linked the microscopic trip data (collected at a frequency of 10 Hz) with a trip-summary file that contains trip-level information, from each instrumented vehicle, and for each trip taken during the study period. The columns in the two files matched well in terms of trip start and end times, vehicle ID and trip ID, distance traveled, average speed, and trip duration. Such concordance increases our confidence in the data.

As stated earlier, the current study focuses on exploring the relationship between driving regimes and most critical correlates i.e. instantaneous driving contexts. This said, descriptive statistics are presented in Table 2 only for instantaneous driving decisions (response variable) and instantaneous driving contexts (explanatory variables) respectively. In Table 2, the explanatory variables are as follow:

1. Objects indicator: 1 if number of objects around host vehicle ≥ 3, 0 otherwise. While we tried different possible categorizations and also used this variable as discrete in the model specifications, the cutoff point of 3 targets provided the most comparable and empirically better (based on AIC) results (Wali et al., 2017). This categorization also helps in comparing the effects of nearby targets on driving regimes across different trips undertaken on different roadway types.
2. Range: indicates the distance of closest object to host vehicle in feet.

### 4.1. Data Aggregation

The SPMD connected vehicle data is collected at a frequency of 10 Hz i.e. 10 BSM packets per second are transmitted between connected vehicles and the infrastructure. This provides the opportunity to conduct microscopic empirical assessment of real-world driving data and vehicular movements that vary substantially over time (Liu and Khattak, 2016). However, as the present study focuses on instantaneous driving decisions, it may be difficult to understand the transition between different regimes, especially within the time frame of one-tenth of a second[17]. Thus, we aggregate the data at relatively lower frequency before conducting detailed econometric analysis of instantaneous driving decisions. However, if the data are aggregated at very lower frequency, it may result in losing short-term extreme or volatile driving decisions (Liu et al., 2015a), which is also a fundamental focus of the present study. According to the study by (Liu et al., 2015a), the feasibility of detecting micro-driving decisions for 1 Hz sampling data (one BSM per second) is 98.54% where 1.46% of the information about micro-decisions can be lost (Liu et al., 2015a). Likewise, if the sampling rate is reduced to 0.5 Hz (one BSM per two seconds), 0.2 Hz (one BSM per five seconds), and 0.1 Hz (one BSM per ten seconds), the information loss can be 4.835%, 17.87%, and 35.86% respectively (Liu et al., 2015a). Given these results and the scope of the present study, we have aggregated the data at 1 Hz (one BSM per second) where averages of the values for each specific variable (identified in

---

[17] We thank the anonymous reviewer for bringing up this conceptual concern to our attention.

Table 1) within one-second are taken[18]. This resulted in a total of 71,434 seconds (i.e. 714,340 BSM packets divided by 10) of real-world connected vehicle driving data.

**(PLACE TABLE 2 ABOUT HERE)**

## 5. Results

### 5.1. Descriptive Statistics

The descriptive statistics presented in Table 2 summarizes each sampled trip by providing mean, standard deviation, minimum and maximum. The distributions of different driving states for each trip e.g. acceleration/deceleration seem reasonable. As compared to mean acceleration/deceleration values, the standard deviation is relatively large for almost all the trips, indicating larger variation in acceleration/deceleration cycles for a given trip. Trips undertaken on freeways (N=12) are relatively longer with a mean and maximum trip duration of 48.6 and 218.4 minutes respectively (Table 2). The trips undertaken on freeways are also observed to be high-speed trips (as compared to those on freeways and state routes) with mean speed of 78.8 mph and maximum mean speed of 81.19 mph respectively. Next, the average trip duration for trips on freeway and state routes (N=4) is 27.8 minutes with maximum trip duration of 34.9 minutes (Table 2). Intuitively, trips on freeways and state routes are also high-speed trips with mean speed of 76.33 mph and maximum mean speed of 89.79 mph respectively (Table 2).

In terms of duration and speed, trips on local roads (N=15) are observed to be both shorter and slower with average trip duration of 13.39 minutes and average speed of 37.98 mph respectively (Table 2). The trips on state and local routes follow similar distribution with mean duration of 29.27 minutes and average speed of 52.94 mph (as compared to average speed of 37.98 mph on local routes) (Table 2). Note that the detailed trip information and the geo-coded trajectories provided in SPMD (Henclewood, 2014)v are not always from start to end of a trip, owing to issues related to privately identifiable data.

To see if the data is characterized by noise, appropriate visualizations are developed. To clarify the relationship between speeds and acceleration, distributions of acceleration are visualized against speed in the top panel of Figure 3. High speeds (>50-55 mph) are associated with smaller acceleration magnitudes as well as smaller dispersion (or volatility) in acceleration/deceleration values. The top right panel in Figure 3 shows the density scatter plot where the bandwidth of acceleration/deceleration values at high speeds is tighter than the bandwidth of acceleration/deceleration values at low speeds. This seems reasonable as vehicle engines should do more work to maintain the same acceleration at higher speeds to overcome increasing air resistance. Therefore, the ability to accelerate a vehicle decreases naturally at higher speeds (Liu and Khattak, 2016, Wang et al., 2015).

To gain further insights regarding data quality, we analyze the distribution of longitudinal vs. lateral accelerations, and the relationship resembles a lozenge shaped distribution which implies that lateral and longitudinal accelerations (or decelerations) do not have large magnitudes simultaneously. Also, the instantaneous driving decisions in longitudinal and lateral directions seem to be inversely correlated with a Pearson correlation coefficient of -0.22, which is in agreement with previous literature (Wang et al., 2015). Such concordance again increases our confidence in the data.

---

[18] Note that we also conducted the entire analyses using original data resolution of 10 Hz. However, doing so did not change our overall inferences regarding the presence and identification of regimes, and its correlations with explanatory variables in typical driving cycle. Results of the analyses conducted at 10 Hz data are available from authors upon request.

**(PLACE FIGURE 3 ABOUT HERE)**

### *5.2. Modeling Results*

Data are used from 38 trips for further analyses[19], the total duration of which is 19.83 hours i.e. it is approximately half of the trip durations for overall 184 trips. As discussed in section 4.1., the data is aggregated at a frequency of 1 Hz (i.e., one BSM per second). Thus, in the present study, the regimes ($s_t$) are same for all 38 trips i.e., regime 1, and regime 2 in two-regime models, and regime 1, regime 2, and regime 3 in three-regime Markov switching models, and that the regimes ($s_t$) can change every one second. For ease of discussion, we first systematically present the results of two-regime dynamic Markov-switching models in section 5.2.1 followed by presenting results for the three-regime dynamic Markov-switching models in section 5.2.2.

### *5.2.1. Two-Regime Dynamic Markov Switching Models:*

We estimated 76 Markov switching regression models to analyze each trip separately, i.e. 38 constant-only instantaneous driving decision models for each trip and 38 instantaneous driving decision models with all explanatory variables. The analyses are conducted as:

- First, to observe the relationships and correlations, for each trip, we estimated simple Ordinary Least Square (OLS) regression models for modeling instantaneous driving decisions (response variable) as a function of number of objects around host vehicle and distance to closest object. Both explanatory variables were statistically significantly associated with modeling instantaneous driving decisions (response variable) at 95% confidence level.
- Second, to capture the evolution of driving behavior followed by time series, constant-only two-regime Markov Switching Dynamic Regression (MSDR) models were estimated for each trip. In the constant-only models, the intercept terms and variances[20] could switch between regimes. In other words, Eq. (4) was estimated. Constant-only models were developed to observe two regimes, regime-dependent means and the associated variances or volatilities. Table 3 illustrates the results of two-regime constant-only models for six trips, whereas Figure 4 illustrates the summary for constant-only models for all 38 trips. In Table 3, the regime-dependent means and variances are reported. Also, mean transition probabilities[21] (1→1, 1→2, 2→1, 2→2) are reported for the selected six trips, where 1→1 can be interpreted as estimated transitional probability of staying in regime 1 in the next period given the driver is observed in regime 1 in current period. Finally, mean durations of each regime are reported in Table 3 and Figure 4.

---

[19] As mentioned earlier, detailed analysis is conducted for 38 trips (out of 43 trips) as 5 trips were excluded from the analysis due to relatively shorter duration (i.e. less than 2 minutes) and no objects around the host vehicle were recorded by Mobile Eye sensor for such trips.

[20] Intuitively, it would be unreasonable to assume that the variance in acceleration be equal to variance in deceleration. Thus, as a first step, two-regime constant-only models were developed with switching intercept-term only. Next, the variances were also allowed to switch between states. Finally, the model with switching intercept and variance term was selected as final model if variance terms were observed to be different in two regimes and statistically significant at 95% confidence level (Hamilton, 1994).

[21] Note that P(1→1) = 1 - P(1→2) and P(2→1) = 1 - P(2→2).

- For ease of discussion, we divide 38 trips (after estimating separate models) into two categories: 1) Category 1: trips on freeways, state routes, and freeway and state routes, and 2) Category 2: trips on local and state routes, and local routes.
- Finally, we estimate full two-regime Markov switching dynamic regression models with full specification as of Equation 6 i.e. $\boldsymbol{\theta} = [\mu_1, \mu_2, \sigma_1{}^2, \sigma_2{}^2, p_{1-2}, p_{2-1}, \beta_{s_t=1}, \beta_{s_t=2}]$. In this model, all estimable parameters $(\mu, \sigma, \beta)$ can switch between the two regimes of a specific driving cycle. Regarding regime-dependent variance term for the full models, we estimated models both with regime-dependent and regime-independent variance terms, and the model that resulted in best fit was selected (discussed later) (Hamilton, 1994). Table 4 illustrates the results of full models (including regime-dependent explanatory variables) for the same trips for which constant-only models are presented in Table 3. Furthermore, Table 5 and 6 summarizes the results of all specified two-regime models, for category 1 and category 2 trips, respectively. For all model parameters as identified in Equation 6, to summarize the distribution of estimated parameters for all trips, Table 5 and 6 present the mean, minimum, and maximum parameter estimates ($\beta$avg, $\beta$min, $\beta$max), standard deviation (Std.dev), and several percentile values (25[th]P, 50[th]P, 75[th]P, and 90[th]P), for category 1 and category 2 trips, respectively.

<div align="center">

**(PLACE TABLE 3 ABOUT HERE)**

**(PLACE FIGURE 4 ABOUT HERE)**

**(PLACE TABLE 4 ABOUT HERE)**

**(PLACE TABLE 5 ABOUT HERE)**

**(PLACE TABLE 6 ABOUT HERE)**

</div>

### 5.2.2. *Three-Regime Dynamic Markov Switching Models:*

As discussed in detail in section 3.1., real-world driving is a complex task and we can anticipate existence of more than two regimes, say three regimes in a typical driving cycle. Thus, it is plausible to also investigate a more generic model specification that may capture common driving regimes, and thus can help in extracting important information related to instantaneous driving decisions embedded in real-world connected vehicle data. Having said this, we estimated 76 three-regime Markov switching regression models to analyze each trip separately, i.e., 38 constant-only three-regime instantaneous driving decision models for each trip and 38 three-regime instantaneous driving decision models with all explanatory variables i.e., full specification. The analyses are conducted as:

- Like the two-regime Markov switching models, for ease of discussion in three-regime specification, we divide the trips (after estimating separate models) into two categories: 1) Category 1: trips on freeways, state routes, and freeway and state routes, and 2) Category 2: trips on local and state routes, and local routes.
- To capture the evolution of driving behavior followed by time series, constant-only three-regime Markov Switching Dynamic Regression (MSDR) models are estimated for each trip. Specifically, the regimes are not observed i.e., we do not a-priori what are the three assumed regimes in a typical driving cycle. Like the two-regime constant only models, the intercept terms and variances can switch between the three regimes. In other words, Equations 7 and 8 are estimated. Constant-only models are developed to observe the three regimes, regime-dependent means and the associated variances or volatilities associated with each regime. Table 7 summarizes the results of three-regime constant-only models for all trips, whereas Figure 3 graphically illustrates the mean intercepts and the associated volatilities associated with each of the three regimes for all the trips. For all model parameters in three-regime constant only models, to summarize the distribution of estimated parameters for all trips, Table 7 also presents the mean, minimum, and maximum parameter estimates (βavg, βmin, βmax), standard deviation (Std.dev), and several percentile values (25thP, 50thP, 75thP, and 90thP), for category 1 and category 2 trips, respectively[22,23].

- Finally, for all the trips, we estimate full three-regime Markov switching dynamic regression models with full specification as of Equation 9 i.e., $\boldsymbol{\theta} = [\mu_1, \mu_2, \mu_3, \gamma_{s_t=1}, \gamma_{s_t=2}, \gamma_{s_t=3}, \sigma_1{}^2, \sigma_2{}^2, \sigma_3{}^2, p_{1-1}, p_{1-2}, p_{2-1}, p_{2-2}, p_{3-1}, p_{3-2}]$. In this model, all estimable parameters $(\mu, \sigma, \beta)$ can switch between the three regimes of a specific driving cycle. Regarding regime-dependent variance term for the full models, we estimated models both with regime-dependent and regime-independent variance terms, and the model that resulted in best fit was selected. Table 8 and 9 summarizes the results of full three-regime models (including regime-dependent explanatory variables) for category 1 and category 2 trips, respectively. Also, Table 8 and 9 present the mean, minimum, and maximum parameter estimates (βavg, βmin, βmax), standard deviation (Std.dev), and several percentile values (25thP, 50thP, 75thP, and 90thP), for three-regime category 1 and category 2 trips, respectively[24]. Also, regime durations and mean transition probabilities[25] are reported (as in Equation 12) for all the trips, where for example, 3→1 can be interpreted as estimated transitional probability of staying in regime 3 in the next period given the driver is observed in regime 1 in current period.

---

[22] The default algorithm we used for maximizing the likelihood functions for all trips in two-regime as well as in three-regime models is modified or quasi Newton-Raphson (NR) algorithm. The three-regime constant only models readily converged for 28 trips, however, for four category 1 trips and six category 2 trips, the three-regime constant only models did not converge. For these trips, we also tried other maximization algorithms such as Berndt-Hall-Hall-Hausman (BHHH), Davidon, Fletcher-Powell (DFP), and Broyden-Fletcher-Goldfarb-Shanno (BFGS) algorithms, however, the models did not converge. The failure of the quasi NR optimization (and other optimization methods) imply that the parameters of the specified three-regime models are not identified by the data, and this is common when attempting to fit a model with too many regimes (Stata, 2016).

[23] As such, 10 trips are dropped from the estimation sample which corresponds to 20,669 seconds (or 0.206 million BSMs) of driving data i.e., 29% of the data in total sample is dropped for the three-regime constant only models.

[24] For fully-specified three-regime models (i.e., including regime dependent explanatory variables), the models did not converge for 14 trips (five category 1 trips and nine category 2 trips). As such, 14 trips are dropped for the estimation sample which corresponds to 14,830 seconds of driving data i.e., 21% of the data in total sample.

[25] Note that P (1→3) = (1 - P(1→1) – P(1→2)), P(2→3) = (1 - P(2→1) – P(2→2)), and P(3→3) = (1 - P(3→1) – P(3→2)).

**(PLACE TABLE 7 ABOUT HERE)**

**(PLACE FIGURE 5 ABOUT HERE)**

**(PLACE TABLE 8 ABOUT HERE)**

**(PLACE TABLE 9 ABOUT HERE)**

## 6. Discussion

In this section, we discuss the results of two-regime and three-regime dynamic Markov switching models. First, the results of two-regime (constant only and models including all explanatory factors- Tables 3 through 6) are discussed followed by a discussion on three-regime Markov switching models (Tables 7 through 9).

### 6.1. Two Regime Dynamic Markov Switching Models

#### 6.1.1. Two-Regime Constant-Only Models (Table 3 and Figure 4)

The constant-only models are developed to investigate whether the volatility of entire driving cycle is sensitive to regimes, i.e. single estimate of variance for the entire driving cycle or is volatility (variance terms) regime dependent? The modeling results (Table 3 and Figure 4) reveal an important finding—that two distinct yet unobserved regimes, acceleration and deceleration, exist and the empirical data strongly favor Markov switching dynamic regression models[26]. Wald tests of linear restrictions were conducted for all 38 constant-only models (for 38 trips), testing the coefficients for intercepts in two regimes for equality (null hypothesis). For all 38 trips, with 99.5% confidence, the null hypothesis was rejected in favor of alternative hypothesis, i.e. the differences in intercept values in two regimes are non-zero (Kodde and Palm, 1986). The existence of two distinct regimes in typical driving cycles (both for category 1 and 2 trips) is shown by the mean positive coefficients for regime 1 (Figure 4), and mean negative coefficient for regime 2 for the same trips (Figure 4). Relevant findings are listed below:

- While Table 3 presents results of Markov switching models for six trips as illustration, similar results were obtained for all 38 sampled trips. By examining the results for all 38 trips in Figure 4, for category 1 trips, the mean acceleration (for all 18 trips) is 0.307 $m/sec^2$ as opposed to mean deceleration of -0.547 $m/sec^2$. Note that for all sampled trips (38 trips), the coefficients for intercept terms were consistently positive and negative, for the two regimes, indicating the existence of two distinct regimes in typical driving cycles. Results show that compared to acceleration, drivers decelerate at a higher rate (intercepts of 0.307 $m/sec^2$ vs -0.547 $m/sec^2$). However, for category 2 trips (Figure 4), the difference between magnitudes of mean acceleration (regime 1) and mean deceleration (regime 2) is relatively large, i.e., mean acceleration (for all 20 trips) is 0.235 $m/sec^2$ whereas mean deceleration is -0.930 $m/sec^2$. This finding indicates that on local routes, drivers may decelerate frequently (and at higher rates) due to presence of traffic controls, i.e., signals, stop signs, and yield signs.

---

[26] Note that the two-regimes were unobserved in the sense that we did not assume a-priori before estimation that acceleration and deceleration are two distinct regimes of a typical driving cycle. Instead, we let the Markov switching framework identify two distinct regimes from data. As an example, some possibilities regarding the two regimes could be, acceleration and deceleration, low and high rate acceleration, low and high rate deceleration, and so on. After estimating the Markov switching models, we eventually learned that acceleration and deceleration are the two typical regimes that characterize a typical driving cycle. Similar to the original Hamilton's Markov switching application to US gross national product data (Hamilton, 1994), we reached this conclusion based on the positive and negative statistically significant intercept terms in the two regimes (Figure 4). However, a positive intercept in regime 1 does not necessary mean that regime 1 is wholly characterized by positive values (i.e., acceleration values). It may be the case that regime 1 (which is identified as acceleration) still contain acceleration values near to zero or negative values near to zero, however, the average intercept term is positive and which makes us conclude that on-overage acceleration is regime 1, and vice versa for regime 2 (i.e., deceleration) (Hamilton, 1994). The concept of unobserved yet distinct regimes will become further clearer in case of three-regime models which are discussed later.

- Importantly, for both category 1 and category 2 trips (Figure 4), deceleration is statistically significantly more volatile than acceleration, noting mean $\sigma_1{}^2$ of 0.224 vs. mean $\sigma_2{}^2$ of 0.301 for category 1 and mean $\sigma_1{}^2$ of 0.373 vs mean $\sigma_2{}^2$ of 0.417 for category 2 (Figure 4). One explanation for this important finding can be that drivers react faster to hazardous or difficult situations, e.g. obstruction or a hard-braking car in front, by decelerating harder as compared to their reaction to more non-hazardous conditions, e.g., an open road with no other vehicles.

- Figure 4 also summarizes the mean duration that driver stays in each regime. For example, for category 1 trips, on average, drivers spend more time accelerating (75 seconds) as compared to decelerating (58 seconds). Finally, referring to Table 3, as expected, it can be observed that both regimes i.e. acceleration and deceleration are highly persistent i.e. mean 1→1 probabilities of 0.91 and 0.88 for category 1 and 2 trips respectively (Table 3).

### 6.1.2. *Two-Regime Specified Models (Table 4 to 6)*

The number of objects and distance to the closest object were added as potential regime-dependent explanatory variables. Like constant-only models, implementation of Markov switching dynamic regression with explanatory variables still support the existence of two distinct regimes. The results in Table 4 suggest that the associations of explanatory variables are significantly different and distinct in two regimes. Drivers respond differently to increasing objects in the acceleration regime as they respond to such a situation during deceleration regime. Wald tests of linear restriction for all 38 trips confirmed this finding[27] (Kodde and Palm, 1986).

Note that a positive sign of the mean parameter estimate in the acceleration regime (Table 4) indicates that an increasing magnitude of acceleration is associated with an increase in explanatory variable, e.g., presence of greater than three objects around the host vehicle. However, a positive sign of the parameter estimate in the deceleration regime indicates decrease in absolute magnitude of deceleration with increase in explanatory variable, e.g., presence of more objects. This association is characterized by a ↓ sign (negative association) in Table 10, which summarizes the associations (accounting for statistical significance at 95% confidence level) of explanatory variables with two regimes for all trips. A negative sign of parameter estimates in Table 4 in the deceleration regime indicates increase in absolute magnitude of deceleration (i.e. a negative value added with negative response value) with unit increase in explanatory variable. This association is conceptualized with ↑ sign (positive association) in Table 10.

---

[27] Wald tests of linear restrictions for all 38 full models were conducted. Specifically, the coefficients for intercepts and $\beta$ for explanatory variables in two regimes were tested for equality (null hypothesis). For all 38 trips, at 99.5% confidence level, the null hypothesis was rejected in favor of alternative hypothesis i.e. the differences in intercept and $\beta$ terms in two regimes are non-zero (Kodde and Palm, 1986).

### 6.1.2.1. *Category 1 trips undertaken on freeways, state, and freeway and state routes*

The results of full models for category 1 trips are summarized in Table 5, while summary of direction of effects for all trips is presented in Table 10. The results suggest that deceleration is high rate regime (as compared to acceleration) with mean intercept estimate of -0.368 $m/sec^2$. Furthermore, similar to the results from constant-only models, deceleration is observed statistically significantly more volatile than acceleration (mean $\sigma_1{}^2$ of 0.203 for acceleration vs mean $\sigma_2{}^2$ of 0.271 for deceleration) (Table 5).

   Turning to the estimation results for category 1 trips (Table 5), in the acceleration regime, on average the number of objects is positively associated with driver propensity to accelerate; note that 50th Percentile $\beta$ is 0.013 in Table 5. Moreover, the association between Objects indicator and acceleration-regime is statistically significantly positive for 8 trips, whereas it is statistically significantly negative for 6 trips[28] (Table 10). The difference in associations of Objects indicator (positive for 44.44% and negative for 33.33% of trips) on driver's propensity to accelerate in the regime 1 may be an outgrowth of drivers having different perceptions regarding their surrounding and thus may make different decisions that match their preferences. However, if a driver is observed to be in the deceleration regime, then the Objects indicator (on average) is negatively associated with driver propensity to decelerate, or indirectly driver is observed to decelerate at a lower rate or even accelerate (i.e. $\beta$avg = 0.076 in Table 5). For the association between Objects indicator and deceleration-regime, it is statistically significantly negative for 11 (61.11%) trips, and positive for only 3 (16.66%) trips, and statistically insignificant for 4 (22.22%) trips (Table 10). Both above findings suggest drivers' tendency (on-average) to get out of crowded situations (characterized by greater than or equal to 3 number of objects around host vehicle) by accelerating (if driver is in acceleration regime) or to decelerate at a lower rate or even accelerate, if a driver is in deceleration regime.

   An increase in distance (in feet) to closest object (Range) is associated with an increase in acceleration, noting that $\beta$avg = 0.065 in the acceleration regime (Table 5). Drivers tend to accelerate when they have more space around them and can freely maneuver their vehicle. Despite the heterogeneity in associations of the Range variable in the deceleration-regime (Li et al., 2017, Ahmed et al., 2017, Khattak et al., 2016, Wali et al., 2017), it is generally statistically significantly negative for 10 trips (55.55%) and positive association for only 5 trips (27.77%) (Table 10).

### 6.1.2.2. *Category 2 trips undertaken on local and state, and local only routes*

Table 6 presents specified models for category 2 trips, and the direction of associations for all trips is presented in Table 10. Similar to category 1 trips where deceleration is the observed high rate regime compared to acceleration, for category 2 trips (i.e. particularly trips on lower functional classification roads), the mean intercepts for acceleration- and deceleration-regime vary significantly i.e., 0.344 $m/sec^2$ vs. -0.776 $m/sec^2$. Likewise, deceleration is more volatile than acceleration as indicated by $\sigma_1^2$ of 0.337 for acceleration and $\sigma_2^2$ of 0.424 for deceleration (Table 6).

---

[28] We remind that results presented throughout hold for the two categories of trips, category 1 and 2, and not for specific roadway types per se. For example, among all the 18 trips in category 1, four trips are undertaken on a mixture of freeway and state routes. Thus, the results presented may not be entirely generalizable for trips on freeways only.

Table 6 shows that parameter estimates for object indicator and range are all significantly different between two regimes for category 2 trips. The magnitudes of differences are reasonable, and partly attributable to the fluctuating traffic conditions due to traffic signals and stop or yield signs on lower classification roads. In the acceleration regime, Object indicator and the Range variable are associated with an increase in acceleration, with $\beta$avg of 0.114 and 0.031 for Objects indicator and Range, respectively (Table 6). The positive associations between objects indicator and acceleration are fairly consistent across sampled trips in sense that for object indicators, the association is positive for 7 (35%) trips and negative for only 2 trips (10%) and statistically insignificant for the rest of the trips (Table 10). The consistent finding for Object indicator is that drivers (on-average) prefer to accelerate given more objects around them on local routes. This finding agrees with the one in category 1 trips, showing that drivers (on-average) tend to get out of crowded situations. For trips on local roads, the finding that increase in Range is associated with drivers' tendency to accelerate is also intuitive, as larger space around the host vehicle will enable drivers to maneuver the vehicles freely. However, this finding is not conclusive in the sense that the association between range and acceleration is positive for 30% of sampled trips whereas it is negative for 25% of the sampled trips, and this requires further investigation.

In the deceleration regime, the Objects indicator is negatively associated with deceleration, i.e. with three or more objects around them, drivers on-average tend to accelerate as indicated by $\beta$avg of 0.019. This finding is again in agreement with the ones observed for category 1 trips. Also, in deceleration regime, the negative association between objects indicator and deceleration holds true for 9 trips while it is positive for only 4 trips (Table 10). Finally, in the deceleration regime, increase in Range is associated with drivers' propensity to accelerate, as expected, and the finding seems conclusive in the sense that drivers in 60% of the sampled trips accelerated with increasing distance to the nearest object (Table 10).

<div align="center">(PLACE TABLE 10 ABOUT HERE)</div>

### 6.2. Three Regime Dynamic Markov Switching Models

#### 6.2.1. Three-Regime Constant-Only Models (Table 7 and Figure 5)

The three-regime constant-only models are developed to identify the three regimes in a typical driving cycle, volatilities associated with each regime, and whether the volatility of entire driving cycle is sensitive to regimes, i.e. single estimate of variance for the entire driving cycle or is volatility (variance terms) regime dependent?

For Category 1 trips, i.e., trips on freeways, state routes, and freeway and state routes, the modeling results (in Table 7 and Figure 5) reveal that the mean intercepts for regime 1, 2, and 3 are 0.542, -0.666, 0.012 respectively. Based on these average intercept values, and its higher magnitudes, regime 1, 2, and 3 can be conceptualized as high rate acceleration, high rate deceleration, and constant/cruise state respectively[29,30]. Moreover, drivers' on-average tend to decelerate at a higher rate than their rate of acceleration ($\beta$avg of -0.666 vs 0.542). Note that for all sampled trips (38 trips), the coefficients for intercept terms were statistically significant, and were consistently positive, negative, and near zero for the three regimes, indicating the existence of three distinct regimes in typical driving cycles. Regarding the volatility associated with each regime in category 1 trips, high rate acceleration is the most volatile ($\sigma_1^2 = 0.501$) followed by high rate deceleration ($\sigma_2^2 = 0.302$) and cruise/constant regime ($\sigma_3^2 = 0.143$). Overall, this finding intuitively suggests that compared to cruise/constant regime, drivers instantaneous driving decisions are more volatile both in "high-rate" acceleration as well as "high-rate" deceleration regime.

For Category 2 trips, i.e., trips on local, local & state routes, the modeling results (in Table 7 and Figure 5) reveal that the mean intercepts for regime 1, 2, and 3 are 0.792, -0.824, 0.014 respectively. Based on these statistics, we identify the three regimes as high-rate acceleration, high-rate deceleration, and cruise/constant regime. Again, and intuitively, drivers tend to decelerate at higher rates than their rates of acceleration (Figure 5). However, in case of category 2 trips, high-rate deceleration ($\sigma_2^2 = 0.580$) is the most volatile regime followed by high-rate acceleration ($\sigma_1^2 = 0.544$), and cruise/constant regime ($\sigma_3^2 = 0.137$). Also, for category 2 trips, the magnitudes of the high rate acceleration and high rate deceleration regimes are higher than the corresponding magnitudes for trips on freeways, state routes, and freeway and state routes (Category 1 trips) (Figure 5). This shows that, given high rate regimes, drivers accelerate and decelerate at higher rates on local roads compared to high rate accelerations and decelerations on freeways.

### 6.2.2. Three-Regime Specified Models (Table 8 and 9)

For the specified three-regime models, number of objects surrounding the host vehicle and distance to the closest object are added as potential regime-dependent explanatory variables. Overall, the results in Table 8 and 9 support the existence of three distinct driving regimes for category 1 and category 2 trips, after controlling for context specific explanatory factors. Like the constant-only three-regime models, the three regimes in specified models can be conceptualized as high-rate acceleration, high-rate deceleration, and constant/cruise regime (Table 7 and 8). Also, the correlations between explanatory factors and instantaneous driving decisions are significantly different and distinct in the three-regime specified models (Table 8 and 9). Wald tests of linear restriction for all the trips confirmed this finding where the coefficients for intercepts and $\beta's$ for explanatory variables in the three regimes were tested for equality (null hypothesis) and the null hypothesis was rejected for all the trips at 99.5% confidence level (Kodde and Palm, 1986).

---

[29] Note that the mean intercept values for acceleration and deceleration (0.542 and -0.666) in three-regime specification are higher than the mean intercept values for acceleration and deceleration (0.307 and -0.547) in two-regime specification.

[30] Like the two-regime specification, the regimes in three-regime specification are unobserved i.e., by simply observing our dependent variable (column vector containing acceleration/deceleration values) directly we cannot know a-priori what the three regimes are. Note that in the two-regime case, it happened to be that by directly observing our response outcome, one could have expected acceleration and deceleration as two regimes. However, in case of three regimes, by visually inspecting the response outcome, it is impossible to infer exactly what the three regimes are and the cut-off points where the regimes change or switch. There can be several possibilities: e.g., 1) cruise state, low rate acceleration, and high rate acceleration, 2) cruise state, low rate deceleration, and high rate deceleration, and so on. It is only after application of Markov-switching models that we can mathematically quantify the three regimes by a data-driven approach, and the average cut-off points associated with each regime from the data at hand. Once the regimes are identified, the correlations between response outcome in each regime and explanatory factors are modeled separately in each regime. The concept of unobserved regimes in Markov switching framework is explicitly explained by (Hamilton, 1994).

Finally, Table 11 summarizes the correlations (accounting for statistical significance at 95% confidence level) between explanatory factors and the instantaneous driving regimes. A positive sign of the mean parameter estimate in the high-rate acceleration regime (Table 8) will indicate drivers' tendency to accelerate (on-average) with an increase in value of explanatory variable. However, a positive sign of the parameter estimate in the high-rate deceleration regime (Table 8) will indicate a decrease in absolute value of deceleration with increase in a value of explanatory value. This association is characterized by a ↓ sign (negative association) in Table 11. Likewise, a negative sign of parameter estimate in the high-rate deceleration regime (Table 8) will indicate an increase in absolute magnitude of deceleration (i.e. a negative value added with negative response value) with unit increase in explanatory variable. This association is thus conceptualized with ↑ sign (positive association) in Table 11.

### 6.2.2.1. *Category 1 trips undertaken on freeways, state, and freeway and state routes*

Before discussing the results of specified three-regime models, we note that five category 1 trips and nine category 2 trips were dropped from the sample due to non-convergence in model estimation. As discussed in section 5.2.2., 21% of the data in total sample is lost. However, for the trips for which the individual models converged, the results provide deeper insights (compared to two-regime models) into the correlation mechanism between instantaneous driving regimes and context-specific situational factors. The key takeaways are:

- The results of specified three-regime models (Table 8) suggest that in high-rate acceleration regime, the number of objects surrounding the host vehicle and the distance to the nearest object on average are negatively correlated with driver's propensity to stay in high-rate acceleration at next instant of time (βavg of -0.257 and -0.121 respectively). This seems intuitive as drivers on average, irrespective of their surroundings, may not stay in high-rate acceleration regime given that they are already in high-rate acceleration regime, and/or the ability to accelerate more at higher rates may be limited. Furthermore, the association between objects indicator and high-rate acceleration regime is statistically significantly negative for 46.2% (as opposed to positive correlation for 15.4% of trips) of the trips. Likewise, the association between range and high-rate acceleration is negative for 53.8% of category 1 trips (compared to only 23.1% of trips where the correlation is positive) (Table 11).

- Likewise, given that driver is in high-rate deceleration regime at current instant of time, the results suggest that with increase in number of objects and distance to the nearest object, drivers on-average are less likely to decelerate further at next instant of time, or drivers indirectly decelerate at a lower rate or can even accelerate at next instant of time. This result is intuitive and is reflected by the positive βavg of 0.058 and 0.013 for object indicator and range respectively (Table 8). Moreover, the relationship between object indicator and high-rate deceleration is negative for 46.2% of the trips (compared to 23.1% of trips with positive correlation), whereas the relationship between range and high-rate deceleration is negative for 61.5% of the trips (compared to only 15.4% of trips with positive association). These findings collectively suggest that in high-rate deceleration regime, drivers (on-average) tend to get out of crowded situations (characterized by greater number of objects around host vehicle) by decelerating at a lower rate or even accelerate at next instant of time.

- Finally, and intuitively, if a driver is in cruise/constant regime at current instant of time, with increasing number of objects around host vehicle and/or with increasing distance to the nearest object s(he) is more likely to accelerate (on average) at next instant of time. Moreover, the statistically significant positive associations between range and constant/cruise regime hold for 61.5% of the sampled trips, compared to only 7.7% of the trips where the correlation between range and cruise/constant regime is negative Table 11).

*6.2.2.2. Category 2 trips undertaken on local, local and state routes:*

- Similar to the results for category 1 trips, the results for category 2 trips suggest that in high rate acceleration regime, increase in both object indicator and range are on-average negatively associated with drivers' tendency to stay in high-rate acceleration regime at next instant of time. As discussed earlier, this may be attributed to the fact that vehicle's ability to accelerate further may be constrained given that vehicle is already in high rate regime.
- For high-rate deceleration regime, our results suggest that increase in number of objects around host vehicle and increase in distance to the nearest object are both negatively associated with drivers' tendency to decelerate further at next instant of time. This is reflected in the average βs of 0.058 and 0.013 for object indicator and range respectively (Table 9). Furthermore, the negative association between object indicator and high-rate deceleration regime holds for 36.3% of the sampled trips whereas the negative association between range and high-rate deceleration regime holds for 54.5% of the sampled trips (Table 11). Note that the associations between the explanatory factors and high-rate deceleration regime are positive only for 9.1% of the sampled trips (Table 11).

**(PLACE TABLE 11 ABOUT HERE)**

### *6.3. Short-Term Regime Predictions*

Markov switching models have a flexible structure for predicting unobserved regimes. Driving regimes can be predicted during each time period (Hamilton, 1993). For details regarding forecasting Markov-switching models by different probability estimation methods, interested readers are referred to (Hamilton, 1993). For demonstration, we use the two-regime model specification for estimating smoothed probabilities that predict the regimes at each time period using all sample data (Hamilton, 1993)(Figure 6). The switching model considers different regime-specific correlations, i.e. instantaneous driving contexts. Figure 3 illustrates the key elements of short-term regime predictions for a 25-minutes trip undertaken on I-94 freeway in Ann Arbor, Michigan. The first panel illustrates the time-series acceleration/deceleration cycle for the entire trip; the second panel illustrates the regime-specific variance; and the last panel illustrates the smoothed probabilities of observing a process in a specific regime at any instant of time. Note that the lowest magnitudes of variance shown in circles correspond to the acceleration regime and vice versa, indicating that deceleration regime is more volatile than acceleration. While the results for all other trips are not presented, they are largely similar.

**(PLACE FIGURE 6 ABOUT HERE)**

## 7. Limitations/Future Work

The study is based on a limited number of trips. It is important to note that this study analyzes micro time-series instantaneous driving decisions during trips, but the application makes it difficult to use the entire large-scale database. Moreover, to extract critical information at the micro-level, each trip should be analyzed separately with two- and three-regime model specifications. Utilization of data from all trips for individual analysis is computationally demanding coupled with the difficulty of interpreting the results in a concise and effective manner. However, once the relationships are established at the microscopic level, it should be easier to predict short-term decisions. Even though the analyzed trips (N = 43) are randomly selected, one-day sample data may not be sufficient for conclusive results. While a two-month SPMD sample dataset is available through the Research Data Exchange (RDE, https://www.its-rde.net/home) website, that data cannot be used due to a substantial number of missing observations about instantaneous driving contexts, i.e. number of objects surrounding host vehicles. Also, the model specification is limited, but it can be enhanced by exploring correlates with other variables when such data become available. Also, we acknowledge that if the "type" of the nearest object could be identified, it could have helped in extracting richer insights. In future, with availability of more detailed data about the type of nearest object, the methodology proposed in this study can be extended to understand how different types of nearest objects may influence the instantaneous driving decisions of host vehicle's driver.

Another important consideration relates to the positions of the vehicles surrounding the host vehicle. Conceptually, both greater number of vehicles around the host vehicle as well as the placement/direction of the vehicles surrounding the host vehicle can influence the drivers' instantaneous driving decisions. To further elaborate the potential influence of vehicles' placement surrounding the host vehicle (social envelope) on the instantaneous driving decisions of the host vehicle, Figure 7 is presented below (Khattak et al., 2015). For details about social interaction and/or gossip algorithms for modeling large-scale behavioral systems, see (Karan and Chakraborty, 2016, Srinivasan et al., 2017, Karan and Chakraborty, 2015). The overall driver behavior estimation can be conceptualized as a Markov Decision Process (MDP) (Khattak et al., 2015). Throughout a typical driving task, the driver is required to optimize his/her policy of instantaneous driving decisions (acceleration/deceleration) based on the number of vehicles surrounding the host vehicle and their placement. For simplicity, assume that the host vehicle is traveling on a three lanes roadway segment. Figure 7a illustrates the time complexity of the driver's policy optimization process. Depending on the number of features (slots around the host vehicle where a vehicle can be present or otherwise) considered, the MDP states grow in the order of $2^n$, where $n$ is the number of features considered. With eight features considered (Figure 7a), the possible number of MDP states are 256. Figure 7c through 7e present few of the possible MDP states. When the host vehicle is surrounded by greater number of vehicles, one can expect that the host driver will accelerate (as our analysis suggests) but only if the slot in front of the host vehicle is empty (Figure 7c and 7d). Contrarily, if the host vehicle is in a situation where the front slot is occupied by another vehicle (Figure 7e), the driver must decelerate no matter he/she is surrounded by greater number of objects or otherwise. Due to the data unavailability about the placements of vehicles surrounding the host vehicle, the driver behavioral models presented in this study cannot capture the influence of "positions" of the surrounding vehicles on the instantaneous driving decisions of the host vehicle. As more detailed data become available in the future, accounting for this dimension in the overall driver behavior estimation can yield in more realistic driver behavioral models in a connected vehicles environment.

<div align="center">(PLACE FIGURE 7 ABOUT HERE)</div>

## 8. Conclusion/Implications

This study focuses on utilizing large-scale high frequency data generated by data acquisition systems (DAS) that are installed in vehicles to facilitate V2V and V2I infrastructure communications via state-of-the-art communication and sensor technologies such as dedicated short-range communications. As part of USDOT Safety Model Pilot Deployment program, real-world large-scale empirical data transmitted between connected vehicles and infrastructure are used to investigate instantaneous driving decisions and its variation with respect to the ecosystem of mapped local traffic states in close proximity surrounding the host vehicle. To achieve the objectives, state-of-the-art time-series methods such as Markov-switching dynamic regression models were applied.

By conducting a detailed analysis of 43 randomly chosen trips that were undertaken on various roadway types, the study explores important questions related to instantaneous driving decisions in connected vehicle environment. Note that, the sampled trips account for 52% of the total one-day sample (714, 340 BSM packets out of total N = 1, 399, 084 BSM packets). To facilitate more meaningful conclusions, the entire vehicle trajectories for 43 randomly selected trips were visualized in Google Earth to identify the roadway functional classification on which the trips were undertaken. As such, significant efforts went into classifying the trips with respect to roadway type, and in processing the large-scale connected vehicle data. Altogether, the 43 trips are undertaken by 34 vehicles whereas few vehicles undertook two or more than two trips. The new proposed methodology helps in understanding instantaneous driving decisions in detail, and for providing answers to the following questions:

- How can driving regimes be characterized in a typical driving cycle?
- What is the level of volatility in each driving regime?
- When do the regimes change or how long do they last?
- Are driver decisions consistent across trips undertaken by different drivers?
- Do correlates vary across the regimes?

To answer the afore-mentioned questions, Expectation Maximization algorithm based on maximum likelihood was used for estimating Markov Switching Dynamic Regression models. First, for simplicity, the study categorized instantaneous short-term driving performance into two unobserved regimes and as such two-regime Markov Switching Dynamic Regression models were estimated for all trips. The results reveal that acceleration and braking are two distinct regimes in a typical driving cycle, with braking showing substantially greater volatility. Compared to braking, acceleration regime typically lasts longer i.e. 75 seconds (switching time on average) for trips on freeways, state routes, and freeway and state routes. In addition, analysis reveals that driver decisions are not consistent across different trips as some drivers show greater volatility than others, especially on local and state, and local roads as expected. Importantly, when more objects surround a vehicle, the tendency is to accelerate even more if a driver is in acceleration regime, and to accelerate or lower the intensity of their braking if driver is in braking regime. Lastly, the magnitudes of associations between key correlates and instantaneous driving behavior vary significantly across the two regimes.

Real-world driving is a complex task and we can anticipate existence of more than two regimes. Thus, we allowed for a more generic dynamic Markov switching model specification where the instantaneous driving decision process was modelled as a three-regime process. The results suggest existence of three distinct and unobserved regimes, which are identified as high-rate acceleration, high-rate deceleration, and cruise/constant regime. Moreover, given in a high-rate regime, drivers on-average tend to decelerate at a higher rate than their rate of acceleration. Importantly, we observed that compared to cruise/constant regime, drivers instantaneous driving decisions are more volatile both in "high-rate" acceleration as well as "high-rate" deceleration regime. Finally, the three-regime specification suggested that in high-rate deceleration regime, drivers (on-average) tend to get out of crowded situations by decelerating at a lower rate or even accelerate at next instant of time.

The results obtained from this study has important implications. First, the study presents an appropriate analytical framework that can help in understanding instantaneous driving decisions and key correlates. Driving decisions primarily depend on surrounding traffic states. An in-depth analysis of such factors is important for understanding driver specific behavior and for developing customized driver based safety applications. For instance, researchers and practitioners can implement the proposed methodology to connected vehicle data generated by specific driver for several trips. For a specific driver, quantification of the associations between instantaneous driving decisions and driving contexts can help us understand driver-specific instantaneous volatility, and to develop hazard anticipation and notification systems if a driver is observed to deviate from his/her normal driving patterns. Furthermore, given a specific driver and keeping in view his/her historical instantaneous driving decisions with respect to local traffic states, alerts and warnings can be provided well in advance to driver specifically if he/she is decelerating. Given that deceleration is consistently observed to be more volatile, such alerts and warnings can potentially help in improving safety and traffic flow disturbances. Finally, an important aspect of developing such hazard anticipation and notification systems is the need to be able to perform short term driving regime predictions. Thus, we demonstrate the potential of dynamic Markov switching models in terms of short-term instantaneous regime prediction at specific instances in time. While the current study focused on instantaneous driving decisions in longitudinal direction only, as part of future work, it would be interesting to develop a methodology for simultaneous analysis of instantaneous driving decisions in longitudinal as well as lateral direction. Such a methodology can potentially help in understanding the correlations between instantaneous driving decisions in longitudinal and lateral directions, and how such decisions can be mapped to surrounding traffic environment.

## 9. Acknowledgement

The research is supported by National Science Foundation (Award number: 1538139). The authors would extend special thanks to the following entities at The University of Tennessee for their support: Transportation Engineering & Science Program and Initiative for Sustainable Mobility. The data used in this study were obtained from Research Data Exchange program maintained by Federal Highway Administration, US DOT. Software Packages R, MATLAB,

Google Earth, and Stata 14.1 are used for data cleaning, linkage, processing, and modeling. The views expressed in this paper are those of the authors, who are responsible for the facts and accuracy of information presented herein. The corresponding author would like to dedicate his part of the effort to his mother, for her mentor-ship and ever-deepening support.

## LIST OF FIGURES

## LIST OF TABLES

**TABLE 1** Variable Descriptions from DAS SPMD, Ann Arbor, Michigan.
Source: SPMD Data Handbook (Henclewood, 2014).

**TABLE 2** Descriptive Statistics of Selected BSM Variables.
Notes: Acceleration/Deceleration are recorded in units of $m/_{s^2}$; range in hundreds of feet; average speed in $miles/_{hr}$; and duration in minutes.

**TABLE 2** Descriptive Statistics of Selected BSM Variables *(Continued)*
Notes:
1. Sample size = 713, 896 BSM records (N = 38 trips)
2. Descriptive statistics for 38 trips are presented as 5 trips were excluded from the analysis due to relatively shorter duration (i.e. less than 2 minutes) and no objects around the host vehicle were recorded by Mobile Eye sensor for such trips.
3. Acceleration/Deceleration are recorded in units of $m/_{s^2}$; range in hundreds of feet; average speed in $m/_{hr}$; and duration in minutes.

**TABLE 3** Two-Regime Constant-Only Markov Switching Regression Models (six selected trips).

**TABLE 4** Two-Regime Full Markov Switching Regression Models (for six selected trips).

**TABLE 5** Summary of specified two-regime models for all trips taken on freeways, state routes, and freeway and state routes (Category 1 trips).
Notes: Objects indicator: 1 if ≥3 number of objects, 0 otherwise; Range: Distance to closest object in hundreds of feet; "Sigma" refers to variance of each regime i.e. $\sigma_1^2$ for regime-acceleration and $\sigma_2^2$ for regime-deceleration. 25P, 50P, 75P, 90P refers to $25^{th}$, $50^{th}$, $75^{th}$, and $90^{th}$ percentile values of estimated parameters for all trips. βavg , βmin , βmax refers to mean, minimum, and maximum parameter estimate for all trips. Std. dev refers to standard deviation of mean parameter estimates (βavg).

**TABLE 6** Summary of specified two-regime models for all trips taken on local and state, and local routes (Category 2 trips)
Notes: Objects indicator: 1 if ≥3 number of objects, 0 otherwise; Range: Distance to closest object in hundreds of feet; "Sigma" refers to variance of each regime i.e. $\sigma_1^2$ for regime-acceleration and $\sigma_2^2$ for regime-deceleration. 25P, 50P, 75P, 90P refers to $25^{th}$, $50^{th}$, $75^{th}$, and $90^{th}$ percentile values of estimated parameters for all trips. βavg , βmin , βmax refers to mean, minimum, and maximum parameter estimate for all trips. Std. dev refers to standard deviation of mean parameter estimates (βavg).

**TABLE 7** Summary of three-regime constant only models for all category 1 and category 2 trips.
**Notes: (a)** Four category 1 and six category 2 trips are dropped due to failure in convergence of three-regime constant only models. See footnote 23 for details.

**TABLE 8** Summary of specified three-regime models for all trips taken on freeways, state routes, and freeway and state routes (Category 1 trips)
Note: (*) Five category 1 trips are dropped due to failure in convergence of three-regime fully specified models. See footnote 18 for details.

**TABLE 9** Summary of specified three-regime models for all trips taken on local and state, and local routes (Category 2 trips)
Note: (*) Nine category 1 trips are dropped due to failure in convergence of three-regime fully specified models. See footnote 18 for details.

**TABLE 10** Two-Regime Markov Switching Models - Summary of direction of effects for all trips.
Note: Row-wise percentages sum up to 100.

**TABLE 11** Three-Regime Markov Switching Models - Summary of direction of effects for all trips

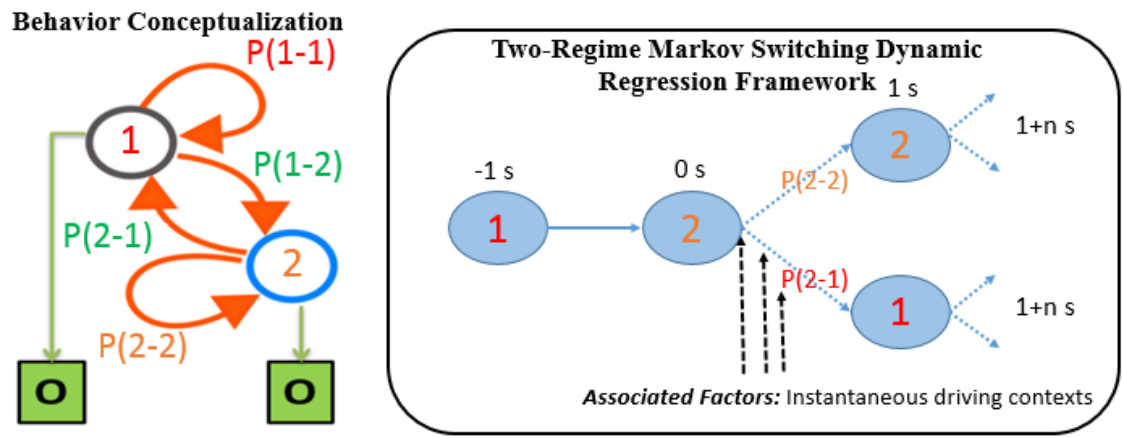

**FIGURE 1** Behavior conceptualization of instantaneous driving decisions in a "two-regime" Markov switching dynamic regression framework (Note: O= Any other unobserved regime).

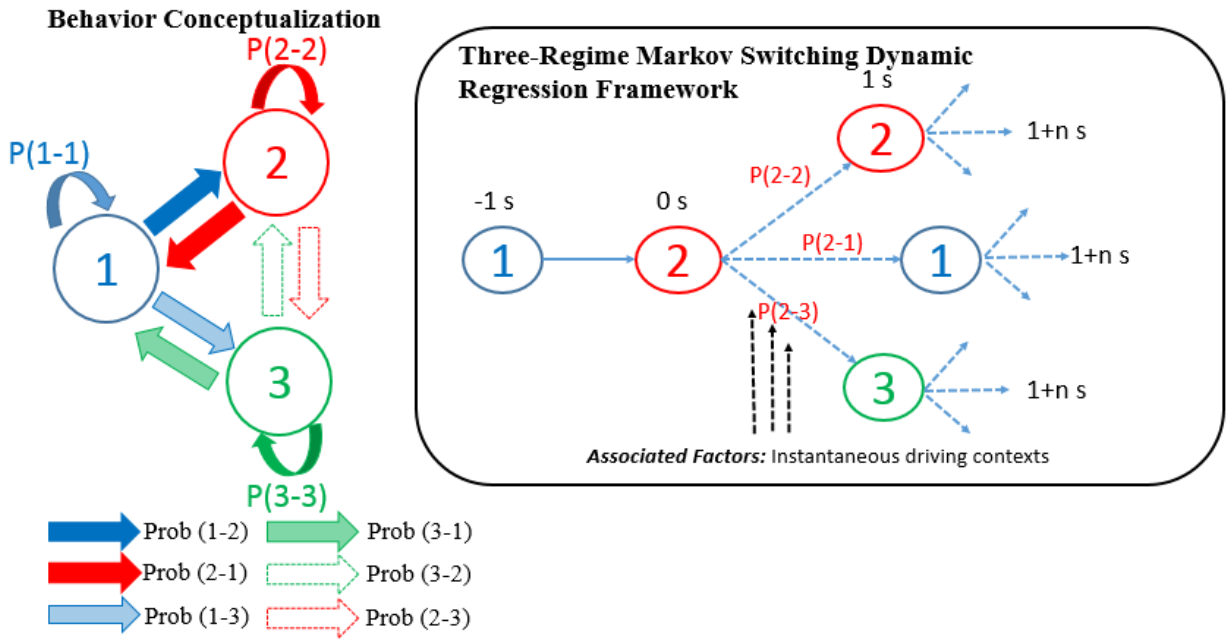

**FIGURE 2** Behavior conceptualization of instantaneous driving decisions in a "three-regime" Markov switching dynamic regression framework.

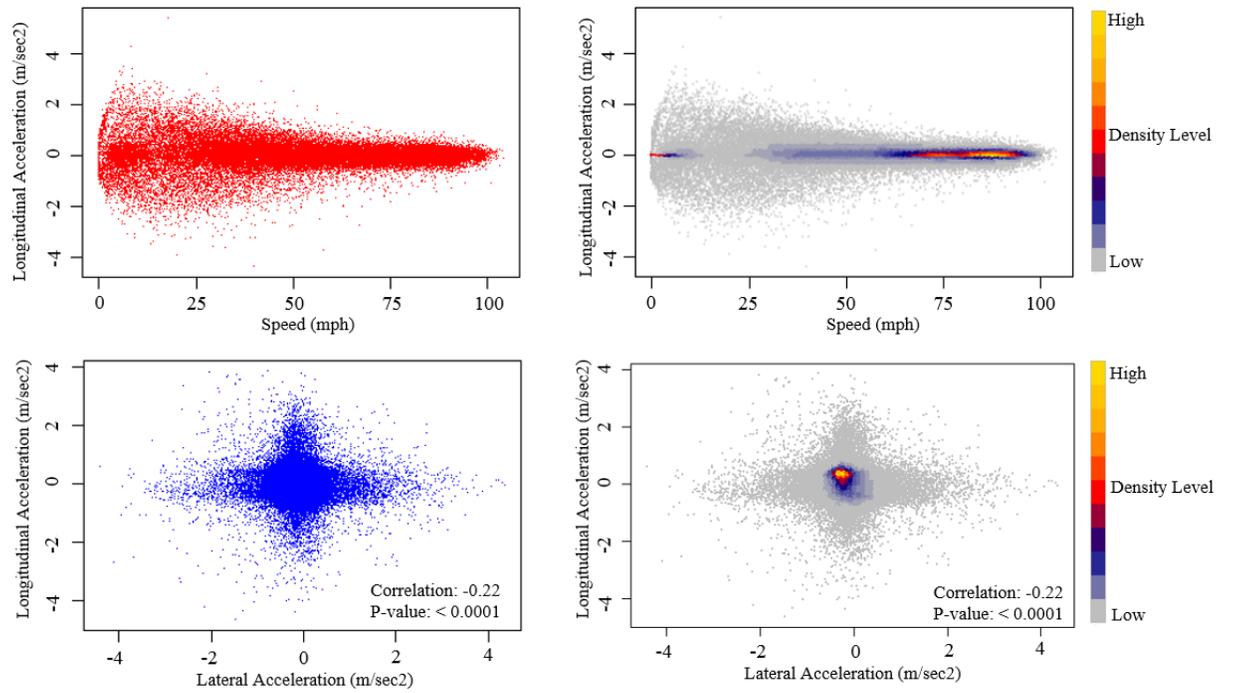

**FIGURE 3** Distributions of speed, longitudinal, and lateral accelerations

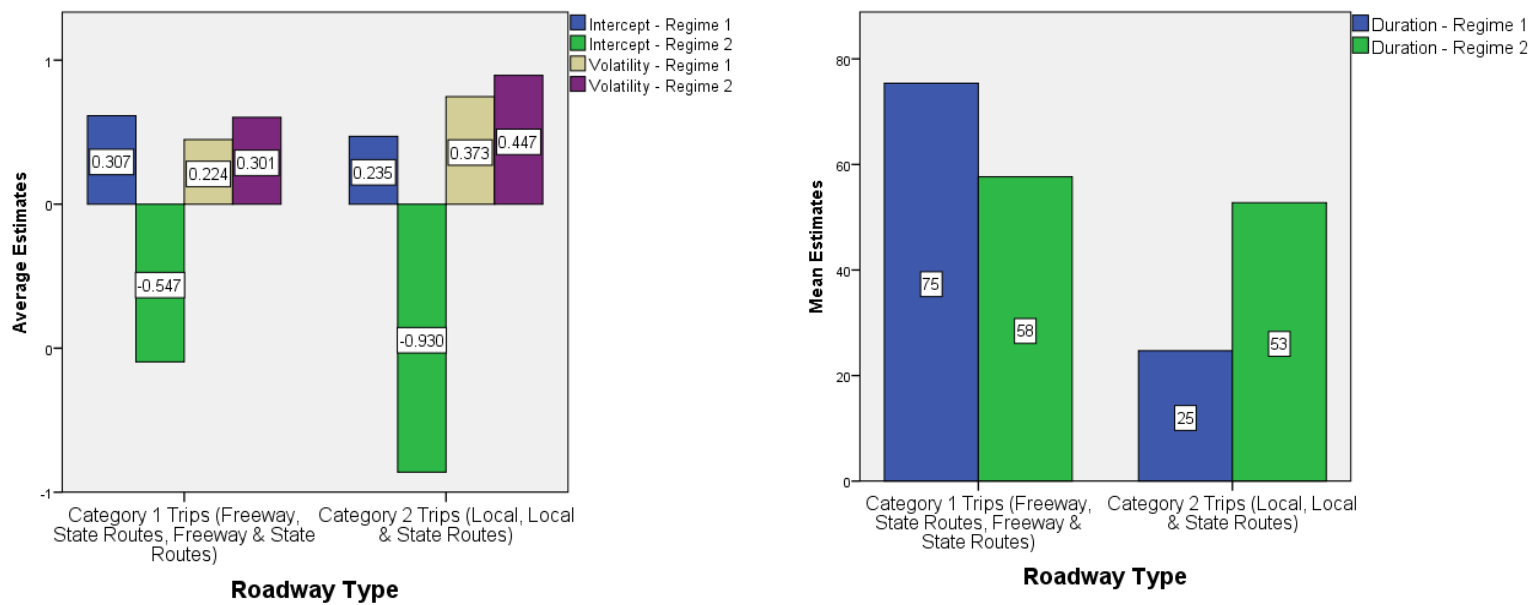

**FIGURE 4** Summary of Two-Regime Constant-Only Markov Switching Regression Models (all 38 trips).

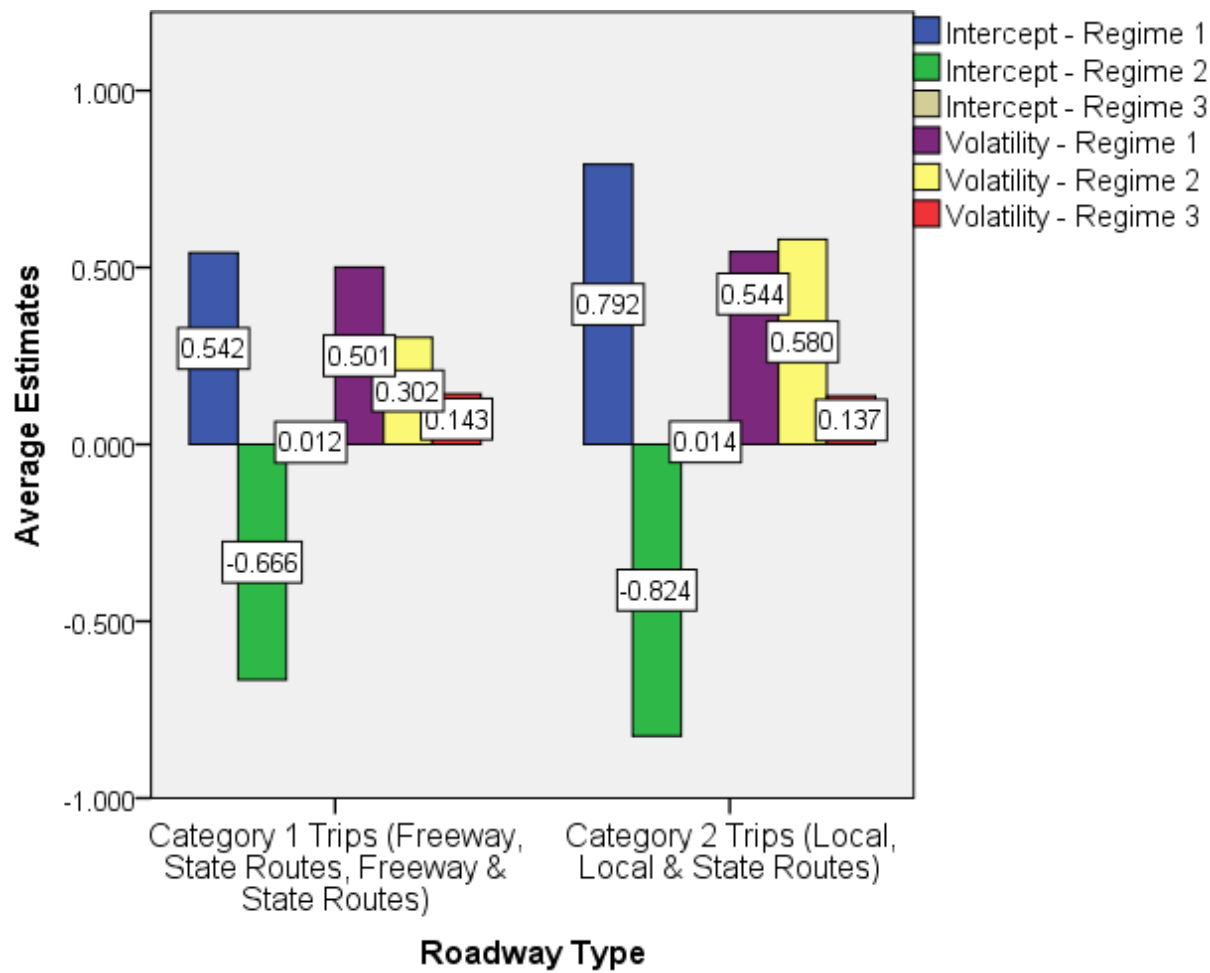

**FIGURE 5** Summary of Three-Regime Constant-Only Markov Switching Regression Models (all 38 trips)

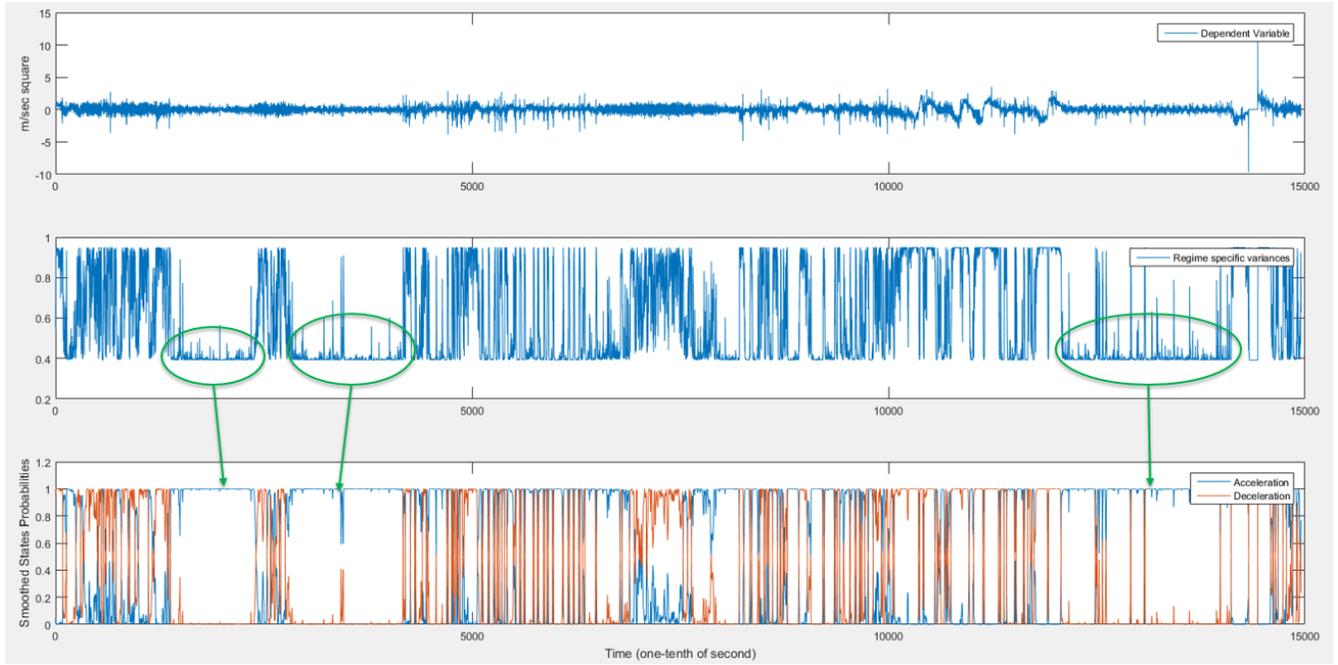

**FIGURE 6** Short term prediction of driving regimes

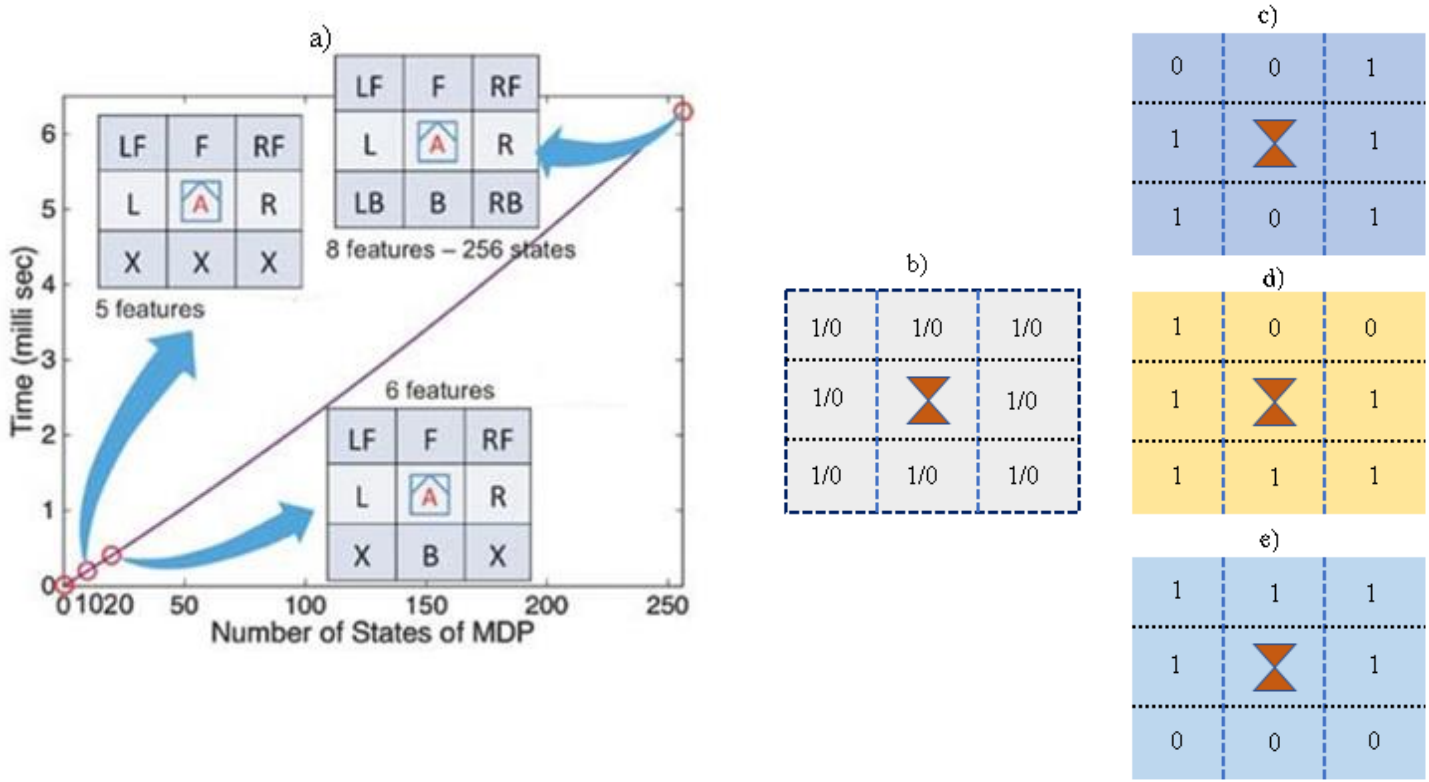

**FIGURE 7:** Illustration of time complexity of the drivers' policy optimization process as a function of number of surrounding vehicles and their placement.

(Notes: MDP is Markov Decision Process; 1/0 indicates if the slot surrounding the host vehicle is occupied or not; L is left; R is right; LF is left front; RF is right front; F is front; LB is left back; RB is right back; B is back; X is slot/feature blocked/not considered.)

**TABLE 1** Variable Descriptions from DAS SPMD, Ann Arbor, Michigan

| Variable | | Description |
|---|---|---|
| **Position** | Altitude | A GPS-based estimate of height above sea level (height above the reference ellipsoid that approximates mean sea level) |
| | Latitude | Current degree of latitude at which the vehicle is located |
| | Longitude | Current degree of longitude at which the vehicle is located |
| **Motion** | Speed (host vehicle) | Current vehicle speed, as determined from the vehicle's transmission |
| | Longitudinal Acceleration | Longitudinal acceleration measured by an Inertial Measurement Unit (IMU) |
| | Lateral Acceleration | Lateral acceleration measured by an IMU |
| **Vehicle Maneuvering** | Accelerator Pedal | Reflects the amount the accelerator pedal is displaced with respect to its neutral position |
| | Brake Pedal | Indicates whether the brake light is on or off |
| | Cruise Control | Indicates whether cruise control is active/engaged |
| | Turn Signal | Provides information regarding the state of the vehicle turn signals |
| **Driving Context** | Number of objects | Number of identified objects, as determined by the Mobileye sensor |
| | Distance to the closest object | Position of the closest object, relative to a reference point on the host vehicle, according to the Mobileye sensor |

Source: SPMD Data Handbook (Henclewood, 2014).

1          **TABLE 2** Descriptive Statistics of Selected BSM Variables

|  | Trip No. | Acc-Dec (Mean/SD/Min/Max) | Number of Targets (Mean/SD/Min/Max) | Range (Mean/SD/Min/Max) | Average Speed | Duration |
|---|---|---|---|---|---|---|
| Freeways & State Routes | 1 | (-0.0043/0.3623/-3.758/2.241) | 0.21/0.40/0/1 | 0.970/0.540/0.093/1.5 | 89.79 | 17.476 |
|  | 2 | (0.008/0.286/-1.5907/2.5759) | 0.56/0.49/0/1 | 0.66/0.47/0.03/1.78 | 68.25 | 26.214 |
|  | 3 | (-0.0112/0.5322/-3.2444/2.3138) | 0.34/0.47/0/1 | 0.29/0.28/0.02/1.5 | 72.48 | 32.768 |
|  | 4 | (0.0031/0.2269/-1.5928/1.3693) | 0.46/0.49/0/1 | 0.99/0.51/0.04/2.24 | 74.82 | 34.953 |
| US State Routes | 1 | (-0.0383/0.3348/-1.6694/2.3972) | 0.09/0.28/0/1 | 0.73/0.40/0.03/1.77 | 53.84 | 4.369 |
|  | 2 | (0.0054/0.6422/-3.60/2.4138) | 0.27/0.44/0/1 | 0.26/0.32/0.02/1.5 | 49.06 | 26.214 |
| Freeways | 1 | (-0.0065/0.4368/-2.921/2.057) | 0.72/0.44/0/1 | 0.35/0.21/0.04/1.59 | 80.48 | 19.661 |
|  | 2 | (-0.0008/0.6055/-2.6779/2.9166) | 0.26/0.44/0/1 | 0.60/0.51/0.02/1.54 | 54.61 | 52.429 |
|  | 3 | (-0.0202/0.6539/-4.37/2.9) | 0.46/0.49/0/1 | 0.40/0.45/0.01/1.77 | 72.02 | 17.476 |
|  | 4 | (-0.0214/0.5873/-1.9694/2.4472) | 0.36/0.48/0/1 | 0.19/0.15/0.02/1.5 | 48.65 | 13.107 |
|  | 5 | (0.0102/0.3439/-1.9032/1.7773) | 0.37/0.48/0/1 | 0.99/0.53/0.03/2.17 | 68.06 | 26.214 |
|  | 6 | (0.0090/0.4562/-1.8164/2.074) | 0.47/0.49/0/1 | 0.80/0.44/0.025/1.78 | 66.88 | 19.661 |
|  | 7 | (0.0018/0.2914/-2.3003/1.8706) | 0.30/0.46/0/1 | 1.05/0.46/0.11/2.16 | 81.19 | 21.845 |
|  | 8 | (0.0166/0.5987/-2.1916/2.6777) | 0.44/0.49/0/1 | 0.23/0.22/0.02/1.5 | 49.2 | 23.815 |
|  | 9 | (-0.0004/0.1863/-1.6645/1.1566) | 0.18/0.39/0/1 | 0.97/0.52/0.02/2.48 | 82.1 | 218.453 |
|  | 10 | (-0.0002/0.1861/-1.8511/1.6970) | 0.18/0.39/0/1 | 1.01/0.47/0.03/2.51 | 76.2 | 196.608 |
|  | 11 | (0.0022/0.4081/-1.9227/2.1484) | 0.47/0.49/0/1 | 0.65/0.47/0.03/1.76 | 72.3 | 26.214 |
|  | 12 | (0.0017/0.3481/-2.2309/1.7925) | 0.30/0.46/0/1 | 0.91/0.53/0.06/ | 72.9 | 21.845 |

2    Notes: Acceleration/Deceleration are recorded in units of $m/s^2$; range in hundreds of feet; average speed in $miles/hr$; and duration in
3    minutes.
4
5
6
7

**TABLE 2** Descriptive Statistics of Selected BSM Variables *(Continued)*

| | Trip No. | Acc-Dec (Mean/SD/Min/Max) | Number of Targets (Mean/SD/Min/Max) | Range (Mean/SD/Min/Max) | Average Speed | Duration |
|---|---|---|---|---|---|---|
| Local Routes | 1 | (0.0064/0.5953/-2.1722/2.0777) | 0.16/0.37/0/1 | 0.71/0.58/0.03/1.68 | 32.23 | 13.107 |
| | 2 | (0.0014/0.900/-3.233/2.4674) | 0.01/0.11/0/1 | 0.28/0.39/0.01/1.5 | 49.69 | 6.554 |
| | 3 | (-0.0028/0.6322/-2.8555/2.8277) | 0.18/0.38/0/1 | 0.43/0.34/0.02/1.5 | 28.19 | 19.661 |
| | 4 | (0.0023/0.6046/-2.100/3.2305) | 0.12/0.33/0/1 | 0.88/0.55/0.02/1.50 | 33.04 | 8.738 |
| | 5 | (0.0241/0.6756/-2.6432/2.0203) | 0.18/0.39/0/1 | 0.27/0.33/0.02/1.5 | 23.73 | 10.923 |
| | 6 | (0.0075/0.4854/-1.911/2.7583) | 0.09/0.29/0/1 | 0.80/0.52/0.04/1.77 | 48.41 | 13.107 |
| | 7 | (-0.0003/0.612/-2.411/2.7886) | 0.17/0.38/0/1 | 0.44/0.45/0.02/1.5 | 39.66 | 15.292 |
| | 8 | (-0.0252/0.5989/-2.3958/1.6883) | 0.24/0.42/0/1 | 0.50/0.53/0.03/1.5 | 33.88 | 4.369 |
| | 9 | (0.0067/0.7795/-3.905/3.480) | 0.05/0.27/0/1 | 0.86/0.57/0.02/1.5 | 40.51 | 34.953 |
| | 10 | (-0.0061/0.5762/-3.1/2.491) | 0.007/0.083/0/1 | 0.93/0.55/0.02/1.5 | 46 | 10.923 |
| | 11 | (0.0105/0.4905/-2.0377/2.0833) | 0.06/0.0818/0/1 | 0.66/0.46/0.03/1.77 | 57.8 | 15.292 |
| | 12 | (0.0173/0.5790/-1.7230/2.6367) | 0.40/0.49/0/1 | 0.53/0.46/0.02/1.5 | 17.5 | 6.554 |
| | 13 | (-0.0062/0.7026/-2.7647/2.5694) | 0.25/0.43/0/1 | 0.51/0.50/0.02/1.52 | 33.1 | 17.476 |
| | 14 | (0.0231/0.485/-1.8722/1.3777) | 0.08/0.27/0/1 | 0.26/0.13/0.04/1.51 | 67.4 | 10.923 |
| | 15 | (0.001/0.6294/-2.4522/2.4110) | 0.04/0.19/0/1 | 0.62/0.51/0.03/2.07 | 18.7 | 13.107 |
| State & local Routes | 1 | (0.0012/0.6672/-2.7908/5.4036) | 0.36/0.48/0/1 | 0.23/0.36/0.01/1.5 | 31.14 | 43.691 |
| | 2 | (0.0128/0.5850/-2.6302/3.3680) | 0.24/0.42/0/1 | 0.46/0.45/0.008/1.63 | 39.85 | 24.030 |
| | 3 | (-0.0067/0.5509/-2.5195/2.4934) | 0.32/0.47/0/1 | 0.48/0.41/0.01/1.62 | 43.97 | 21.845 |
| | 4 | (-0.0045/0.5982/-3.0861/3.1) | 0.02/0.16/0/1 | 1.25/0.43/0.03/1.78 | 76.74 | 32.768 |
| | 5 | (0.0070/0.2776/-1.3715/1.7664) | 0.07/0.0266/0/1 | 1.32/0.35/0.14/1.53 | 73 | 24.030 |

Notes:

1. Sample size = 713, 896 BSM records (N = 38 trips)

2. Descriptive statistics for 38 trips are presented as 5 trips were excluded from the analysis due to relatively shorter duration (i.e. less than 2 minutes) and no objects around the host vehicle were recorded by Mobile Eye sensor for such trips.

3. Acceleration/Deceleration are recorded in units of $m/s^2$; range in hundreds of feet; average speed in $m/hr$; and duration in minutes.

**TABLE 3** Two-Regime Constant-Only Markov Switching Regression Models (six selected trips)

| Constant-only Models | | Freeway | Freeway | Freeway & State Route | Freeway & State Route | Local Route | Local Route |
|---|---|---|---|---|---|---|---|
| Acceleration-Regime 1 | $\beta$ | 0.149 | 0.104 | 1.1297 | 0.147 | 0.2568 | 0.141 |
| | z-score | 16 | 5.23 | 26.01 | 12.61 | 6.01 | 7.93 |
| Deceleration-Regime 2 | $\beta$ | -1.019 | -1.104 | -0.0163 | -0.739 | -1.412 | -1.194 |
| | z-score | -24.86 | 5.23 | -2.69 | -17.04 | -11.19 | -15.5 |
| Regime 1 - Variance Parameter | $\beta$ | 0.4422 | 0.467 | 0.11635 | 0.377 | 0.6646 | 0.453 |
| | Std. Error | 0.0063 | 0.014 | 0.0021 | 0.007 | 0.028 | 0.012 |
| Regime 2 - Variance Parameter | $\beta$ | 0.5811 | 0.341 | 0.1163 | 0.5323 | 0.6891 | 0.443 |
| | Std. Error | 0.0225 | 0.035 | 0.0021 | 0.0222 | 0.0715 | 0.05 |
| Transition prob: 1→2 | $\beta$ | 0.017 | 0.0144 | 0.0926 | 0.0284 | 0.0215 | 0.062 |
| | Std. Error | 0.0027 | 0.005 | 0.0506 | 0.0048 | 0.009 | 0.004 |
| Transition prob: 2→2 | $\beta$ | 0.8834 | 0.873 | 0.9979 | 0.8682 | 0.8763 | 0.866 |
| | Std. Error | 0.0179 | 0.04 | 0.0011 | 0.0213 | 0.0473 | 0.037 |
| Expected Duration: Regime 1 | $\beta$ | 58.58 | 69.12 | 10.7922 | 35.096 | 46.505 | 61.38 |
| | 95% Conf. Interval | 43.0,79.9 | 34.6,139.0 | 4.0,32.8 | 25.1,49.0 | 20.0,109.5 | 34.4,110.1 |
| Expected Duration: Regime 2 | $\beta$ | 8.582 | 7.891 | 71.31 | 7.589 | 8.08 | 7.464 |
| | 95% Conf. Interval | 6.3,11.6 | 4.3,15.2 | 59.1,81.1 | 5.5,10.4 | 4.0,17.7 | 4.4,13.2 |

**TABLE 4** Two-Regime Full Markov Switching Regression Models (for six selected trips)

| Full Models | | Freeway | Freeway | Freeway & State Route | Freeway & State Route | Local Route | Local Route |
|---|---|---|---|---|---|---|---|
| Acceleration-Regime 1 | Constant (std. error) | 0.0666 (0.0159) | 0.077 (0.031) | 1.47 (0.082) | 0.161 (0.019) | 0.273 (0.051) | 0.1172 (0.025) |
| | Objects indicator (std. error) | 0.146 (0.020) | -0.083 (0.035) | -0.328 (0.086) | 0.030 (0.015) | -0.320 (0.322) | 0.229 (0.056) |
| | Range (std. error) | 0.032 (0.017) | 0.279 (0.131) | -0.349 (0.106) | -0.138 (0.037) | -0.084 (0.158) | 0.041 (0.021) |
| Deceleration-Regime 2 | Constant (std. error) | -1.494 (0.073) | -1.002 (0.094) | -0.051 (0.0124) | -0.806 (0.057) | -1.603 (0.180) | -1.183 (0.096) |
| | Objects indicator (std. error) | 0.513 (0.080) | -0.063 (0.102) | 0.0154 (0.006) | 0.090 (0.051) | 1.3707 (0.614) | 0.559 (0.102) |
| | Range (std. error) | 0.162 (0.073) | -0.546 (0.592) | 0.041 (0.012) | -0.258 (0.099) | 0.227 (0.097) | -0.169 (0.129) |
| Regime 1 - Variance Parameter | $\beta$ | 0.2265 | 0.466 | 0.229 | 0.402 | 0.665 | 0.401 |
| | Std. Error | 0.003 | 0.014 | 0.004 | 0.007 | 0.072 | 0.011 |
| Regime 2 - Variance Parameter | $\beta$ | 0.2265 | 0.338 | 0.1145 | 0.201 | 0.655 | 0.456 |
| | Std. Error | 0.003 | 0.035 | 0.002 | 0.0035 | 0.072 | 0.033 |
| Transition prob: 1→2 | $\beta$ | 0.015 | 0.013 | 0.098 | 0.026 | 0.0206 | 0.017 |
| | Std. Error | 0.002 | 0.005 | 0.053 | 0.004 | 0.008 | 0.005 |
| Transition prob: 2→2 | $\beta$ | 0.846 | 0.878 | 0.997 | 0.845 | 0.8755 | 0.899 |
| | Std. Error | 0.022 | 0.04 | 0.001 | 0.024 | 0.048 | 0.027 |

**TABLE 5** Summary of specified two-regime models for all trips taken on freeways, state routes, and freeway and state routes (Category 1 trips)

| | Variable | βavg | Std.dev | βmin | βmax | 25P | 50P | 75P | 90P |
|---|---|---|---|---|---|---|---|---|---|
| Acceleration - Regime 1 | Constant | 0.366 | 0.457 | 0.013 | 1.475 | 0.055 | 0.129 | 0.518 | 1.240 |
| | Objects indicator | -0.003 | 0.164 | -0.328 | 0.460 | -0.078 | 0.013 | 0.062 | 0.146 |
| | Range | 0.065 | 0.218 | -0.330 | 0.608 | -0.026 | 0.017 | 0.216 | 0.279 |
| | Duration-Acc | 49.659 | 38.157 | 10.180 | 150.713 | 19.100 | 40.060 | 72.494 | 105.312 |
| | Sigma-Acc | 0.203 | 0.119 | 0.074 | 0.466 | 0.121 | 0.150 | 0.248 | 0.461 |
| Deceleration - Regime 2 | Constant | -0.568 | 0.486 | -1.494 | -0.052 | -0.994 | -0.451 | -0.109 | -0.071 |
| | Objects indicator | 0.076 | 0.237 | -0.272 | 0.548 | -0.059 | 0.034 | 0.090 | 0.538 |
| | Range | 0.095 | 0.287 | -0.546 | 0.665 | 0.025 | 0.064 | 0.162 | 0.658 |
| | Duration-Dec | 80.900 | 123.787 | 5.040 | 462.590 | 7.119 | 11.768 | 152.030 | 245.900 |
| | Sigma-Dec | 0.271 | 0.244 | 0.074 | 1.109 | 0.123 | 0.178 | 0.347 | 0.445 |
| Transition Probabilities | 1→1 | 0.918 | 0.065 | 0.802 | 0.998 | 0.859 | 0.915 | 0.993 | 0.995 |
| | 2→1 | 0.039 | 0.029 | 0.007 | 0.098 | 0.013 | 0.028 | 0.056 | 0.085 |

Notes: Objects indicator: 1 if ≥3 number of objects, 0 otherwise; Range: Distance to closest object in hundreds of feet; "Sigma" refers to variance of each regime i.e. $\sigma_1^2$ for regime-acceleration and $\sigma_2^2$ for regime-deceleration. 25P, 50P, 75P, 90P refers to 25th, 50th, 75th, and 90th percentile values of estimated parameters for all trips. βavg , βmin , βmax refers to mean, minimum, and maximum parameter estimate for all trips. Std. dev refers to standard deviation of mean parameter estimates (βavg).

**TABLE 6** Summary of specified two-regime models for all trips taken on local and state, and local routes (Category 2 trips)

|  | Variable | βavg | Std.dev | βmin | βmax | 25P | 50P | 75P | 90P |
|---|---|---|---|---|---|---|---|---|---|
| Acceleration - Regime 1 | Constant | 0.344 | 0.441 | -0.006 | 1.692 | 0.113 | 0.174 | 0.333 | 1.092 |
|  | Objects indicator | 0.114 | 0.621 | -1.519 | 2.122 | -0.033 | 0.058 | 0.267 | 0.419 |
|  | Range | 0.031 | 0.312 | -0.814 | 0.701 | -0.100 | 0.017 | 0.102 | 0.460 |
|  | Duration-Acc | 43.443 | 33.034 | 5.558 | 146.365 | 19.763 | 38.103 | 54.402 | 80.490 |
|  | Sigma-Acc | 0.337 | 0.147 | 0.112 | 0.615 | 0.216 | 0.333 | 0.429 | 0.551 |
| Deceleration - Regime 2 | Constant | -0.776 | 0.509 | -1.604 | -0.102 | -1.189 | -0.802 | -0.228 | -0.138 |
|  | Objects indicator | 0.019 | 0.675 | -2.167 | 1.371 | -0.133 | 0.117 | 0.322 | 0.575 |
|  | Range | 0.109 | 0.465 | -1.077 | 1.028 | -0.154 | 0.102 | 0.328 | 0.697 |
|  | Duration-Dec | 29.255 | 42.197 | 5.472 | 147.324 | 8.061 | 10.342 | 23.319 | 110.604 |
|  | Sigma-Dec | 0.424 | 0.248 | 0.112 | 1.018 | 0.226 | 0.401 | 0.538 | 0.812 |
| Transition Probabilities | 1→1 | 0.912 | 0.053 | 0.817 | 0.993 | 0.876 | 0.903 | 0.955 | 0.991 |
|  | 2→1 | 0.043 | 0.043 | 0.007 | 0.180 | 0.018 | 0.026 | 0.055 | 0.097 |

Notes: Objects indicator: 1 if ≥3 number of objects, 0 otherwise; Range: Distance to closest object in hundreds of feet; "Sigma" refers to variance of each regime i.e. $\sigma_1^2$ for regime-acceleration and $\sigma_2^2$ for regime-deceleration. 25P, 50P, 75P, 90P refers to 25th, 50th, 75th, and 90th percentile values of estimated parameters for all trips. βavg , βmin , βmax refers to mean, minimum, and maximum parameter estimate for all trips. Std. dev refers to standard deviation of mean parameter estimates (βavg).

**TABLE 7** Summary of three-regime constant only models for all category 1 and category 2 trips.

| | Regimes | Parameters | βavg | Std.dev | βmin | βmax | 25P | 50P | 75P | 90P |
|---|---|---|---|---|---|---|---|---|---|---|
| Freeway, State Routes, Freeway & State Routes (N = 14).[a] | High Rate Acc - Regime 1 | Intercept | 0.542 | 0.336 | 0.065 | 0.965 | 0.113 | 0.639 | 0.829 | 0.953 |
| | | Sigma (Volatility) | 0.501 | 0.316 | 0.049 | 1.245 | 0.336 | 0.454 | 0.672 | 1.081 |
| | | Duration | 10.006 | 3.503 | 5.903 | 16.589 | 6.889 | 8.833 | 12.913 | 15.772 |
| | High Rate Dec - Regime 2 | Intercept | -0.666 | 0.414 | -1.576 | -0.129 | -0.936 | -0.646 | -0.264 | -0.138 |
| | | Sigma (Volatility) | 0.302 | 0.218 | 0.049 | 0.690 | 0.107 | 0.282 | 0.475 | 0.635 |
| | | Duration | 7.651 | 3.241 | 3.749 | 15.489 | 4.987 | 7.143 | 8.866 | 14.020 |
| | Cruise - Regime 3 | Intercept | 0.012 | 0.034 | -0.059 | 0.059 | -0.014 | 0.022 | 0.035 | 0.057 |
| | | Sigma (Volatility) | 0.143 | 0.080 | 0.049 | 0.260 | 0.066 | 0.131 | 0.227 | 0.254 |
| | | Duration | 33.991 | 35.978 | 9.422 | 138.165 | 15.717 | 18.400 | 41.081 | 110.321 |
| Local, State and Local Routes (N = 14).[a] | High Rate Acc - Regime 1 | Intercept | 0.792 | 0.186 | 0.507 | 1.237 | 0.679 | 0.785 | 0.867 | 1.108 |
| | | Sigma (Volatility) | 0.544 | 0.092 | 0.433 | 0.794 | 0.480 | 0.526 | 0.593 | 0.713 |
| | | Duration | 9.566 | 2.575 | 5.946 | 13.139 | 7.464 | 8.940 | 12.495 | 12.908 |
| | High Rate Dec - Regime 2 | Intercept | -0.824 | 0.216 | -1.310 | -0.586 | -0.945 | -0.807 | -0.624 | -0.587 |
| | | Sigma (Volatility) | 0.580 | 0.096 | 0.398 | 0.790 | 0.532 | 0.555 | 0.650 | 0.749 |
| | | Duration | 9.050 | 2.081 | 5.364 | 12.895 | 7.147 | 9.594 | 10.588 | 12.132 |
| | Cruise - Regime 3 | Intercept | 0.014 | 0.013 | -0.019 | 0.038 | 0.007 | 0.012 | 0.021 | 0.033 |
| | | Sigma (Volatility) | 0.137 | 0.061 | 0.056 | 0.312 | 0.106 | 0.127 | 0.160 | 0.245 |
| | | Duration | 21.280 | 14.680 | 6.472 | 57.200 | 11.603 | 14.377 | 34.665 | 47.258 |

**Notes:** (a) Four category 1 and six category 2 trips are dropped due to failure in convergence of three-regime constant only models. See footnote 23 for details.

**TABLE 8** Summary of specified three-regime models for all trips taken on freeways, state routes, and freeway and state routes (Category 1 trips)

| | Regimes | Parameters | βavg | Std.dev | βmin | βmax | 25P | 50P | 75P | 90P |
|---|---|---|---|---|---|---|---|---|---|---|
| Freeway, State Routes, Freeway & State Routes (N = 14). * | High Rate Acc - Regime 1 | Intercept | 0.693 | 0.393 | 0.040 | 1.239 | 0.383 | 0.788 | 0.943 | 1.200 |
| | | # of objects | -0.277 | 0.659 | -2.359 | 0.184 | -0.242 | -0.075 | 0.011 | 0.155 |
| | | Range | -0.104 | 0.278 | -0.549 | 0.423 | -0.334 | -0.126 | 0.027 | 0.382 |
| | | Sigma (Volatility) | 0.330 | 0.202 | 0.061 | 0.676 | 0.109 | 0.345 | 0.512 | 0.638 |
| | | Duration | 10.643 | 4.041 | 5.930 | 18.784 | 7.563 | 9.972 | 13.496 | 17.837 |
| | High Rate Dec - Regime 2 | Intercept | -0.811 | 0.569 | -2.326 | -0.147 | -1.128 | -0.686 | -0.429 | -0.162 |
| | | # of objects | 0.164 | 0.450 | -0.272 | 1.561 | -0.018 | 0.055 | 0.216 | 1.050 |
| | | Range | 0.206 | 0.534 | -0.767 | 1.575 | -0.080 | 0.188 | 0.430 | 1.160 |
| | | Sigma (Volatility) | 0.324 | 0.178 | 0.061 | 0.578 | 0.124 | 0.387 | 0.450 | 0.560 |
| | | Duration | 8.142 | 1.950 | 4.983 | 11.877 | 7.108 | 7.806 | 9.168 | 11.667 |
| | Cruise - Regime 3 | Intercept | -0.011 | 0.043 | -0.070 | 0.078 | -0.049 | -0.001 | 0.019 | 0.057 |
| | | # of objects | -0.001 | 0.050 | -0.144 | 0.054 | -0.010 | 0.002 | 0.030 | 0.053 |
| | | Range | 0.084 | 0.202 | -0.116 | 0.694 | -0.006 | 0.018 | 0.089 | 0.517 |
| | | Sigma (Volatility) | 0.231 | 0.351 | 0.049 | 1.322 | 0.068 | 0.134 | 0.235 | 0.997 |
| | | Duration | 37.969 | 36.668 | 8.152 | 128.455 | 16.128 | 17.742 | 59.744 | 113.238 |
| | Transition Probabilities | 1→1 | 0.029 | 0.039 | 0.000 | 0.126 | 0.002 | 0.019 | 0.028 | 0.114 |
| | | 1→2 | 0.212 | 0.310 | 0.030 | 0.939 | 0.063 | 0.100 | 0.134 | 0.913 |
| | | 2→1 | 0.810 | 0.220 | 0.084 | 0.915 | 0.842 | 0.871 | 0.890 | 0.913 |
| | | 2→2 | 0.096 | 0.040 | 0.002 | 0.144 | 0.070 | 0.106 | 0.129 | 0.139 |
| | | 3→1 | 0.090 | 0.250 | 0.003 | 0.920 | 0.010 | 0.025 | 0.034 | 0.567 |
| | | 3→2 | 0.823 | 0.335 | 0.046 | 0.992 | 0.938 | 0.944 | 0.983 | 0.991 |

Note: (*) Five category 1 trips are dropped due to failure in convergence of three-regime fully specified models. See footnote 24 for details.

**TABLE 9** Summary of specified three-regime models for all trips taken on local and state, and local routes (Category 2 trips)

| | Regimes | Parameters | βavg | Std.dev | βmin | βmax | 25P | 50P | 75P | 90P |
|---|---|---|---|---|---|---|---|---|---|---|
| Local, State and Local Routes (N = 14). * | High Rate Acc - Regime 1 | Intercept | 0.764 | 0.156 | 0.499 | 0.948 | 0.612 | 0.852 | 0.887 | 0.936 |
| | | # of objects | -0.259 | 0.690 | -1.859 | 0.251 | -0.226 | 0.005 | 0.150 | 0.249 |
| | | Range | -0.121 | 0.265 | -0.825 | 0.085 | -0.188 | -0.025 | 0.056 | 0.084 |
| | | Sigma (Volatility) | 0.503 | 0.125 | 0.280 | 0.792 | 0.459 | 0.486 | 0.552 | 0.754 |
| | | Duration | 9.756 | 2.874 | 5.588 | 13.237 | 7.111 | 8.985 | 12.654 | 13.186 |
| | High Rate Dec - Regime 2 | Intercept | -0.771 | 0.273 | -1.039 | -0.166 | -1.018 | -0.789 | -0.650 | -0.223 |
| | | # of objects | 0.058 | 0.210 | -0.221 | 0.403 | -0.167 | 0.078 | 0.204 | 0.379 |
| | | Range | 0.013 | 0.450 | -1.031 | 0.808 | -0.183 | 0.059 | 0.194 | 0.721 |
| | | Sigma (Volatility) | 0.552 | 0.098 | 0.393 | 0.787 | 0.497 | 0.538 | 0.561 | 0.758 |
| | | Duration | 9.345 | 1.721 | 6.876 | 11.947 | 7.842 | 9.964 | 10.727 | 11.863 |
| | Cruise - Regime 3 | Intercept | 0.020 | 0.052 | -0.006 | 0.176 | -0.001 | 0.003 | 0.018 | 0.145 |
| | | # of objects | -0.079 | 0.212 | -0.697 | 0.072 | -0.092 | -0.005 | 0.010 | 0.063 |
| | | Range | -0.022 | 0.107 | -0.335 | 0.067 | -0.014 | 0.005 | 0.019 | 0.061 |
| | | Sigma (Volatility) | 0.111 | 0.034 | 0.054 | 0.158 | 0.095 | 0.111 | 0.142 | 0.155 |
| | | Duration | 17.606 | 10.541 | 7.322 | 37.475 | 10.235 | 14.665 | 24.588 | 37.220 |
| | Transition Probabilities | 1→1 | 0.021 | 0.024 | 0.000 | 0.080 | 0.000 | 0.020 | 0.029 | 0.071 |
| | | 1→2 | 0.089 | 0.033 | 0.033 | 0.150 | 0.076 | 0.083 | 0.114 | 0.144 |
| | | 2→1 | 0.889 | 0.021 | 0.855 | 0.916 | 0.872 | 0.900 | 0.906 | 0.916 |
| | | 2→2 | 0.062 | 0.019 | 0.032 | 0.086 | 0.038 | 0.063 | 0.078 | 0.086 |
| | | 3→1 | 0.042 | 0.019 | 0.021 | 0.087 | 0.024 | 0.040 | 0.051 | 0.081 |
| | | 3→2 | 0.926 | 0.035 | 0.863 | 0.973 | 0.902 | 0.932 | 0.959 | 0.973 |

Note: (*) Nine category 1 trips are dropped due to failure in convergence of three-regime fully specified models. See footnote 24 for details.

**TABLE 10** Two-Regime Markov Switching Models - Summary of direction of effects for all trips

| Roadway Type | Driving Regimes | Variable | ↑ | ↓ | Not Significant at 95% CL |
|---|---|---|---|---|---|
| Freeways & State Routes (N = 18 trips) | Acceleration | Constant | 18 *(100%)* | 0 *(0%)* | 0 *(0%)* |
| | | Objects indicator | 8 *(44.44%)* | 6 *(33.33%)* | 4 *(22.22%)* |
| | | Range | 10 *(55.55%)* | 5 *(27.77%)* | 3 *(16.66%)* |
| | Deceleration | Constant | 0 *(0%)* | 18 *(100%)* | 0 *(0%)* |
| | | Objects indicator | 3 *(16.66%)* | 11 *(61.11%)* | 4 *(22.22%)* |
| | | Range | 2 *(11.11%)* | 11 *(61.11%)* | 5 *(27.77%)* |
| Local, State & Local Routes (N = 20 trips) | Acceleration | Constant | 19 *(95%)* | 0 *(0%)* | 1 *(5%)* |
| | | Objects indicator | 7 *(35%)* | 2 *(10%)* | 11 *(55%)* |
| | | Range | 6 *(30%)* | 5 *(25%)* | 9 *(45%)* |
| | Deceleration | Constant | 0 *(0%)* | 20 *(100%)* | 0 *(0%)* |
| | | Objects indicator | 4 *(20%)* | 9 *(45%)* | 7 *(35%)* |
| | | Range | 5 *(25%)* | 12 *(60%)* | 3 *(15%)* |

Note: Row-wise percentages sum up to 100.

**TABLE 11** Three-Regime Markov Switching Models - Summary of direction of effects for all trips

| Roadway Type | Driving Regimes | Variable | ↑ | ↓ | Not Significant at 95% CL |
|---|---|---|---|---|---|
| Freeways & State Routes (N = 13 trips) | High Rate Acceleration- Regime 1 | Constant | 13 *(100%)* | 0 | 0 |
| | | Objects indicator | 2 *(15.4%)* | 6 *(46.2%)* | 5 *(38.5%)* |
| | | Range | 3 *(23.1%)* | 7 *(53.8%)* | 3 *(23.1%)* |
| | High Rate Deceleration - Regime 2 | Constant | 0 | 13 *(100%)* | 0 |
| | | Objects indicator | 3 *(23.1%)* | 6 *(46.2%)* | 4 *(30.8%)* |
| | | Range | 2 *(15.4%)* | 8 *(61.5%)* | 3 *(23.1%)* |
| | Constant/Cruise around 0 - Regime 3 | Constant | 3 *(23.1%)* | 6 *(46.2%)* | 4 *(30.8%)* |
| | | Objects indicator | 4 *(30.8%)* | 2 *(15.4%)* | 7 *(53.8%)* |
| | | Range | 8 *(61.5%)* | 1 *(7.7%)* | 4 *(30.8%)* |
| **Roadway Type** | **Driving Regimes** | **Variable** | **↑** | **↓** | **Not Significant at 95% CL** |
| Local, Local & State Routes (N = 11 trips) | High Rate Acceleration- Regime 1 | Constant | 11 (100%) | 0 | 0 |
| | | Objects indicator | 2 *(18.2%)* | 4 *(36.4%)* | 5 *(45.5%)* |
| | | Range | 1 *(9.1%)* | 4 *(36.34%)* | 6 *(54.5%)* |
| | High Rate Deceleration - Regime 2 | Constant | 0 | 11 *(100%)* | 0 |
| | | Objects indicator | 1 *(9.1%)* | 4 *(36.3%)* | 6 *(54.5%)* |
| | | Range | 1 *(9.1%)* | 6 *(54.5%)* | 4 *(36.4%)* |
| | Constant/Cruise around 0 - Regime 3 | Constant | 1 *(9.1%)* | 1 *(9.1%)* | 9 *(81.8%)* |
| | | Objects indicator | 3 *(27.2%)* | 4 *(36.3%)* | 4 *(36.3%)* |
| | | Range | 3 *(27.2%)* | 2 *(18.1%)* | 6 *(54.5%)* |